\documentclass[fleqn,usenatbib]{mnras}

\usepackage{lscape}
\usepackage{fancyhdr}
\usepackage{newtxtext,newtxmath}

\usepackage{datetime} 
\usepackage[utf8]{inputenc}

\usepackage{graphicx}	
\usepackage{amsmath}	
\usepackage{amssymb}	
\usepackage{comment}
\usepackage{bm}
\usepackage{subcaption}

\usepackage{relsize}
\usepackage[T1]{fontenc}
\pdfoutput=1
\DeclareRobustCommand{\VAN}[3]{#2}
\let\VANthebibliography\thebibliography
\def\thebibliography{\DeclareRobustCommand{\VAN}[3]{##3}\VANthebibliography}

\title[Three-point intrinsic alignments of halos]{Three-point intrinsic alignments of dark matter halos in the IllustrisTNG simulation}
\author[S. Pyne et al.]{
Susan Pyne,$^{1}$\thanks{E-mail: ucapsep@ucl.ac.uk } Ananth Tenneti$^{1}$ 
       and Benjamin Joachimi$^{1}$
\\
$^{1}$Department of Physics and Astronomy, University College London, Gower Street, London WC1E 6BT, UK\\
}

\begin{document}
\label{firstpage}
\pagerange{\pageref{firstpage}--\pageref{lastpage}}
\maketitle

\begin{abstract}

We use the IllustrisTNG suite of cosmological simulations to measure intrinsic alignment (IA) bispectra of dark matter subhalos between redshifts 0 and 1. We decompose the intrinsic shear field into E- and B-modes and find that the bispectra $B_{\delta\delta\mathrm{E}}$ and $B_{\delta\mathrm{EE}}$, between the matter overdensity field, $\delta$, and the E-mode field, are detected with high significance.  We also model the IA bispectra analytically using a method consistent with the two-point non-linear alignment model.  We use this model and the simulation measurements to infer the intrinsic alignment amplitude $A_\mathrm{IA}$ and find that values of $A_\mathrm{IA}$ obtained from IA power spectra and bispectra agree well at scales up to \mbox{$k_\mathrm{max}= 2  \, h \mathrm{Mpc}^{-1}$}. For example at $z=1$   $A_\mathrm{IA} =  2.13 \pm$ 0.02 from the cross power spectrum between the matter overdensity and E-mode fields and  $A_\mathrm{IA} =2.11 \pm$ 0.03 from $B_{\delta\delta \mathrm{E}}$. This demonstrates that a single physically motivated model can  jointly model two-point and three-point statistics of intrinsic alignments, thus enabling a cleaner separation between intrinsic alignments and cosmological weak lensing signals.
\end{abstract}
\section{Introduction}

Gravitational tidal effects cause dark matter halos and the galaxies within them to preferentially align with the cosmological large-scale structure. Thus the shapes and orientations of galaxies and halos may be correlated across cosmological distances.  This phenomenon is known as intrinsic alignment (IA)   \citep{kirk2012cosmological,joachimi2015galaxy,kiessling2015galaxy,kirk2015galaxy,troxel2015intrinsic}. 

Intrinsic alignments  have been widely studied, mainly because the IA effects mimic cosmic shear and thus are an undesirable contaminant of weak lensing shape measurements. Controlling this systematic uncertainty has been a major concern for recent weak lensing surveys, such as the Kilo-Degree Survey (KiDS-1000)\footnote{http://kids.strw.leidenuniv.nl/index.php}  \citep{joachimi2021kids} and the Dark Energy Survey\footnote{https://www.darkenergysurvey.org} \citep{secco2022dark}.  For forthcoming surveys such as  \textit{Euclid}\footnote{http://sci.esa.int/euclid/} \citep{laureijs2011euclid} and the Vera C. Rubin Observatory Legacy Survey of Space and Time\footnote{https://www.lsst.org}, control of systematics will become even more pressing. These surveys will measure the shapes of billions of galaxies, reducing statistical uncertainties, so that uncontrolled systematics will be a limiting issue.  Looking further ahead, as understanding of intrinsic alignments develops it is also possible to see this signal as a valuable cosmological probe in its own right  which future surveys will be able to exploit \citep{chisari2013cosmological,taruya2020improving}.  All these considerations suggest that it is worthwhile to explore intrinsic alignments as broadly as possible.

So far most theoretical and observational studies of intrinsic alignments have used two-point statistics; relatively few  have taken the further step of considering three-point measurements.  However there is evidence that intrinsic alignments affect two-point and three-point weak lensing statistics differently \citep{huterer2006systematic,semboloni2008sources,troxel2011self, troxel2012self,pyne2021self}.  For this reason most  work which has investigated three-point statistics aimed to use them  to self-calibrate intrinsic alignment contamination in weak lensing data.

Three-point IA statistics have already been successfully measured from both survey data and simulations. 
On the observational side \citet{semboloni2010weak} measured three-point aperture mass statistics from the Cosmic Evolution Survey \citep{scoville2007cosmic} and  \citet{fu2014cfhtlens} built on this work  using  the larger Canada France Hawaii Telescope Lensing Survey (CFHTLenS; \citealt{heymans2012cfhtlens}). Both these studies  showed that using three-point statistics could help improve constraints on cosmological parameters.  Early simulation results were described by  \citet{semboloni2008sources} who measured three-point intrinsic alignment aperture mass statistics from simulations described in \citet{heymans2006potential}. The main focus of this work was a comparison between the amplitudes of intrinsic alignment and weak lensing statistics. 

Another strand of work has involved analytical modelling of three-point IA statistics, building on methods developed for two-point statistics, in particular the non-linear alignment model (NLA) \citep{hirata2004intrinsic,bridle2007dark}. This  postulates that the intrinsic alignment of halos is related to the tidal gravitational field at an earlier redshift.  This methodology can been extended to three-point statistics in a natural but empirical way \citep{ troxel2012self,merkel2014theoretical,pyne2021self}.    Other intrinsic alignment models, for  example the halo-model of \citet{schneider2010halo} and tidal alignment models such as  \citet{blazek2015tidal}, have generally only been implemented for two-point statistics.  An exception is the development by  \citet{vlah2020eft,vlah2021galaxy}  of effective field theory models of galaxy alignments. These authors modelled two-point statistics at next-to-leading order and three-point statistics at leading order.

 Despite the body of work described above, an analytical model for three-point intrinsic alignment statistics has never been tested against simulations.   In this work we aim to fill this gap.  As a starting point we consider dark matter subhalos rather than galaxies. Since the intrinsic alignment of galaxies is related to that of their host halos, our work can be seen as a first step towards a model for galaxies.

Our approach to  measurement from simulations is based on that of  \citet{kurita2021power} who developed a novel method to measure three-dimensional intrinsic alignment power spectra of dark matter halos which they applied to the high-resolution DarkQuest suite of simulations \citep{nishimichi2019dark,miyatake2021cosmological}. This has a resolution of 2048$^3$ particles in a periodic cubic box size of \mbox{1 $h^{-1}$ Gpc}. 
Using the same methodology, we measure all theoretically non-zero intrinsic alignment bispectra for equilateral and specific isosceles triangles from the IllustrisTNG\footnote{https://www.tng-project.org} simulation suite.  

\citet{kurita2021power} compared their simulation results with the two-point non-linear alignment model and concluded that at linear scales the model matches the simulation results, with a scale-independent intrinsic alignment amplitude, but that at 
non-linear scales the model breaks down and the amplitude is no longer constant.
We similarly compare our bispectrum measurements with an analytical model based on perturbation theory which is in keeping with the two-point NLA  model.

In Section~\ref{sect:sims} we describe the simulation suite and the methods used to measure dark matter subhalo shapes. 
Section~\ref{sect:IA} explains how we measure IA power spectra and bispectra from simulations, and Section~\ref{sect:LAM} discusses the NLA model and our analytical model for three-point IA statistics.
In Section~\ref{sect:results} we present our IA power spectrum and bispectrum measurements from simulations and compare these with analytical results, and in Section~\ref{sect:comp} we discuss how these compare with previous work. We summarise and discuss  possible further work in Section~\ref{sect:discussion}.  Appendices discuss other details, including signal-to-noise ratios and a possible phenomenological modification to the non-linear alignment model.

\section{Simulations}\label{sect:sims}
\subsection{Characteristics of the simulations}\label{sect:simchars}
We use the publicly available cosmological  simulation suite IllustrisTNG
\citep{marinacci2018first,naiman2018first,nelson2018first,nelson2019illustristng,pillepich2018first,springel2018first}. 
Each simulation in the suite uses the moving-mesh code \textsc{Arepo} \citep{springel2010pur}\footnote{https://arepo-code.org} and self-consistently solves for the coupled evolution of dark matter, cosmic gas, luminous stars and supermassive black holes from the starting redshift \mbox{$z=127$} to the present day, based on a cosmology consistent with \textit{Planck}  \citep{ade2016planck}: $\Omega_\Lambda=0.6911, \Omega_\mathrm{m}=0.3089,\Omega_\mathrm{b}=0.0486, 
 \sigma_8 = 0.8159, n_\mathrm{s} =0.9667, h =0.6774$.  
 Specifically we use the IllustrisTNG300-1 hydrodynamic simulation which  has the largest simulation box size within the suite of simulations available from the public data release \citep{nelson2019illustristng}.  The simulation box volume is 300 Mpc$^3$ (comoving)  with \mbox{$2\,500^3$} dark matter particles and an equal initial number of gas cells.

The publicly available data includes catalogues of halos identified using the Friends-of-Friends algorithm \citep{davis1985evolution}. The  \textsc{subfind} algorithm \citep{springel2001populating} is used to identify substructures within the halos.  This  algorithm defines the centre of the subhalo  as the location of the most bound particle, and provides the positions of all dark matter particles in the halo.  

For our analysis we select dark matter subhalos  from the simulations at three redshifts, $z=0.0, 0.5, 1.0$. The subhalo mass is  the total sum of the masses of all individual particles belonging to a given subhalo as identified by the \textsc{subfind} algorithm. The maximum subhalo masses in the simulation are approximately $3.4 \times 10^{14}$, $7.4 \times 10^{14}$ and $1.2 \times 10^{15}$ $\mathrm{M}_\odot$ at $z=1.0, 0.5, 0$ respectively.  We select subhalos within the mass range $4 \times 10^{10}$ to $10^{14} \; h^{-1} \mathrm{M}_\odot$ for our analysis.

\subsection{Measurement of 3D halo shapes from simulations}

We start by measuring 3D ellipsoidal subhalo shapes. For this we adopt the widely used method based on the inertia tensor $I_{ij}$. This is defined as
\begin{align}
I_{ij} = \frac{\sum_n m_n x_{ni}x_{nj}}{\sum_n{m_n}}\ ,\label{eq:IT}
\end{align}
where $m_n$ is the mass of the $n$th particle and $x_{ni}, x_{nj}$ are its position coordinates relative to the centre of the halo.  The semi-axes $a, b, c$ of the ellipsoid  are obtained from the eigenvalues $\lambda_a, \lambda_b,\lambda_c$ of the inertia tensor,  with $a =\sqrt{\lambda_a}$ and so on. We set $c\le b\le a$  and define axis ratios $s=c/a$ and $q=b/a$. The eigenvectors of the inertia tensor determine the orientation of the axes.

In order to obtain well-resolved shapes, we choose only subhalos with a minimum of 1000 dark matter particles, which is consistent with the particle number threshold adopted in earlier studies, for example \citet{tenneti2015intrinsic}. With this threshold the minimum subhalo mass is about $4 \times 10 ^{10} \; h^{-1} \mathrm{M}_\odot$ .

As an improvement on equation (\ref{eq:IT}) we use the reduced inertia tensor \citep{tenneti2015intrinsic} 
which gives more weight to particles which are closer to the centre of a subhalo, thus avoiding potential problems with defining its outer edge. The reduced inertia tensor is defined as 
\begin{align}
\tilde{I}_{ij} = \frac{\sum_n m_n\frac{x_{ni}x_{nj}}{r_n^2}}{\sum_n{m_n}}\ ,
\end{align}
where
\begin{align}
r_n^2 = \sum_{i=1}^3 x_{ni}^2\ .
\end{align}

Rather than taking the \lq raw\rq \ axis ratios defined above we use the iterative approach described in \citet{tenneti2015intrinsic} to recover the shape of an isodensity surface \citep{schneider2012shapes}.   In this method  the principal axes of the ellipsoids are rescaled iteratively while the enclosed volume is kept constant. After each rescaling, particles outside the ellipsoidal volume are discarded.  The process is repeated until  the fractional change in the axis ratios is below a predefined limit, in our case 1 per cent.

\section{Intrinsic alignment spectra from simulations}\label{sect:IA}
\subsection{Ellipticity and tidal shear}

The ellipticity,  $\epsilon$, of a subhalo shape is a  spin-2 quantity  (it is invariant under rotations of integer multiples of $\pi$).  It can be  parametrized in several ways in terms of the shape and orientation of the subhalo. All parametrizations are essentially equivalent so we can make a choice  which suits the problem at hand.  
 In many cases it is convenient to express the ellipticity as \mbox{$\epsilon = \epsilon_+ +\mathrm{i}  \epsilon_\times$} where $ \epsilon_+$ represents stretching along a defined axis and $\epsilon_\times$ represent stretching along an axis at 45\degr  \ to this.   It is always possible to find such a decomposition \citep{stebbins1996weak}. 
 
  We can then express the  two components of the ellipticity as
\begin{align}
  \epsilon_+&= \bigg(\frac{a-b}{a+b}\bigg)\cos2\theta\label{eq:eps+}\\
   \epsilon_\times&= \bigg(\frac{a-b}{a+b}\bigg)\sin2\theta\label{eq:epsx}\ ,
 \end{align}
where $a$ and $b$ are the semi-major and semi-minor axes of the ellipse. Here we take the line of sight to be the  $z$-axis and consider  ellipses projected on to the plane perpendicular to this \citep{kurita2021power,shi2021power}.   The angle $\theta$ is an arbitrary choice which we take to be the position angle of the major axis with respect to the $x$-axis in the $x$-$y$ plane, as determined by the relevant eigenvector. Equations~(\ref{eq:eps+}) and (\ref{eq:epsx}) clearly satisfy the requirements for a spin-2 quantity.  Moreover, under a parity transformation $\epsilon_+$ is unchanged but $\epsilon_\times \rightarrow -\epsilon_\times$ \citep{schneider2002b}.

This definition of $\epsilon$, in terms of \mbox{$(a-b)/(a+b)$}, is commonly used for weak lensing shear because it provides an unbiased estimator of the shear and does not depend on the ellipticity distribution of source galaxies \citep{seitz1997steps,viola2014probability}.   Even though we are considering intrinsic ellipticity rather than cosmic shear we adopt this definition for consistency.  It has the added advantage that the tidal shear $\gamma=\gamma_+ +\mathrm{i}\gamma_\times$ can be assumed to be directly proportional to the ellipticity. 

Other studies of intrinsic alignments, for example  \citet{blazek2015tidal} and \citet{shi2021power}, used an alternative definition of ellipticity which replaces    \mbox{$(a-b)/(a+b)$} with \mbox{$(a^2-b^2)/(a^2+b^2)$} in equations~(\ref{eq:eps+}) and (\ref{eq:epsx}). This version is more readily comparable with observations but does not provide an unbiased shear estimator \citep{schneider1995}.  Moreover rather than the shear being directly related to the ellipticity,  an extra responsivity factor \mbox{$\mathcal{R}=1-\epsilon_\mathrm{rms}^2$}  is required \citep{bernstein2002shapes}, where \mbox{$\epsilon_\mathrm{rms}= \sqrt{\langle \epsilon_+^2\rangle}=\sqrt{\langle \epsilon_\times^2\rangle}$}. The responsivity  measures the average response of $ \epsilon_{+,\times}$ to  $ \gamma_{+,\times}$ so that 
\mbox{$ \gamma_{+,\times} = \epsilon_{+,\times}/2\mathcal{R}$}. 
With our definition of the ellipticity we do not need to consider the responsivity and can assume that $\epsilon$ directly traces the tidal shear field $\bm{\gamma}$. 
Thus  $\gamma_+$ and $\gamma_\times$ are given by  equations~(\ref{eq:eps+}) and (\ref{eq:epsx}).  
\subsection{Decomposition into E- and B-modes}
For studies of intrinsic alignments it is convenient to go a step further and  decompose the shear field into a curl-free (E-mode) component $\gamma_\mathrm{E}$ and a gradient-free (B-mode) component  $\gamma_\mathrm{B}$  \citep{kamionkowski1998theory,crittenden2002discriminating}.  We define these by the equations
 \begin{align}
 \nabla^2 \gamma_\mathrm{E}(\bm{x}) &=(\partial_{x}\partial_{x}-\partial_{y}\partial_{y})\gamma_+(\bm{x}) +2 \partial_{x}\partial_{y}\gamma_\times(\bm{x})\label{eq:del2gammaE}\\
 \nabla^2 \gamma_\mathrm{B}(\bm{x}) &=(\partial_{x}\partial_{x}-\partial_{y}\partial_{y})\gamma_\times(\bm{x}) -2 \partial_{x}\partial_{y}\gamma_+(\bm{x})\ ,\label{eq:del2gammaB}
 \end{align} 
 where $\bm{x} $ is the configuration space position and $\partial_{x}\partial_{x} \equiv \partial^2/\partial x^2$, and so on.

This decomposition takes a simpler form in Fourier space. We define Fourier space  coordinates to be \mbox{$\bm{k}=(k_x,k_y,k_z)$} and choose the $k_z$-axis to be along the line of sight.  
The derivatives in equations~(\ref{eq:del2gammaE}) and  (\ref{eq:del2gammaB}) change to multiplicative factors so we get
 \begin{align}
k_{xy}^2 \tilde{\gamma}_\mathrm{E}(\bm{k})&= (k_x^2-k_y^2)\tilde{\gamma}_+(\bm{k})+2k_xk_y\tilde{\gamma}_\times(\bm{k})\\
k_{xy}^2 \tilde{\gamma}_\mathrm{B}(\bm{k})&= (k_x^2-k_y^2)\tilde{\gamma}_\times(\bm{k})-2k_xk_y\tilde{\gamma}_+(\bm{k})\ ,
 \end{align}
 where $k_{xy}^2 = k_x^2 + k_y^2$ and $\tilde{\gamma}_+$ and $\tilde{\gamma}_\times$ are the Fourier transforms of  equations~(\ref{eq:eps+}) and (\ref{eq:epsx}). 
 Alternatively these expressions can be written in terms of the angle $\phi = \tan^{-1}(k_x/k_y)$ giving \citep{kurita2021power,shi2021power} 
 \begin{align}
\tilde{\gamma}_\mathrm{E}(\bm{k})&=\tilde{\gamma}_+(\bm{k})\cos{2\phi} + 
 \tilde{\gamma}_\times(\bm{k})\sin{2\phi} \label{eq:gammaE}\\
 \tilde{\gamma}_\mathrm{B}(\bm{k})&=  \tilde{\gamma}_\times(\bm{k})\cos{2\phi} -\tilde{\gamma}_+(\bm{k})\sin{2\phi}\ .\label{eq:gammaB}
 \end{align}
Even though  $\tilde{\gamma}_\mathrm{E}$ and $\tilde{\gamma}_\mathrm{B}$ are measured in the plane perpendicular to the line of sight, the wavevector  $\bm{k}$ is three-dimensional: \mbox{$\bm{k}= k(\sqrt{1-\mu^2 }\cos{\phi},\sqrt{1-\mu^2} \sin{\phi},\mu)$}, where   $\cos^{-1}(\mu)$ is the angle between $\bm{k}$ and the $k_z$-axis \citep{blazek2015tidal,kurita2021power}.

\subsection{Intrinsic alignment power spectra and bispectra}

Equations~(\ref{eq:gammaE}) and  (\ref{eq:gammaB}) lead directly to the intrinsic alignment power spectrum between  the Fourier transforms of the E-mode shear, $\tilde{\gamma}_\mathrm{E}$, and matter density contrast, $\tilde{\delta}$, and also the auto power spectra of $\tilde{\gamma}_\mathrm{E}$ and $\tilde{\gamma}_\mathrm{B}$:
\begin{align}
\langle\tilde{\delta}(\bm{k})\tilde{\gamma}_\mathrm{E}(\bm{k}^\prime)\rangle&= (2\mathrm{\pi})^3\delta_\mathrm{D}^3(\bm{k}+\bm{k}^\prime)P_{\delta\mathrm{E}}(\bm{k})\label{eq:PSDE}\\
\langle\tilde{\gamma}_\mathrm{E}(\bm{k})\tilde{\gamma}_\mathrm{E}(\bm{k^\prime})\rangle&= (2\mathrm{\pi})^3\delta_\mathrm{D}^3(\bm{k}+\bm{k}^\prime)P_{\mathrm{E}\mathrm{E}}(\bm{k})\label{eq:PSEE}\\
\langle\tilde{\gamma}_\mathrm{B}(\bm{k})\tilde{\gamma}_\mathrm{B}(\bm{k^\prime})\rangle&= (2\mathrm{\pi})^3\delta_\mathrm{D}^3(\bm{k}+\bm{k}^\prime)P_{\mathrm{B}\mathrm{B}}(\bm{k})\label{eq:PSBB}\ ,
\end{align}
where $\langle\rangle$ denotes the ensemble average and $\delta_\mathrm{D}^3$ is the three-dimensional Dirac delta function.
From parity considerations these are the only possible non-zero intrinsic shear power spectra \citep{stebbins1996weak, kamionkowski1998theory,crittenden2002discriminating,schneider2002b}. Power spectra involving a single B-mode shear will switch sign if 
$\bm{k}\rightarrow -\bm{k}$ which is physically impossible unless the  spectra are zero.  Theoretically the B-mode auto power spectrum also vanishes to first order but in practice it may be non-zero because of Poisson shot noise due to the finite sampling of the halo positions, which is also present in the E-mode auto power spectrum. (See \citet{blazek2019beyond} and \citet{kurita2021power} for more detailed discussions of shape noise).

The formalism of equations~(\ref{eq:PSDE}) to (\ref{eq:PSBB}) can be extended to intrinsic alignment bispectra. Again, any bispectrum which includes an odd number of B-modes  can be expected to be zero by parity arguments   so there are five possible non-zero bispectra:

\begin{align}
\langle\tilde{\delta}(\bm{k}_1)\tilde{\delta}(\bm{k}_2)\tilde{\gamma}_\mathrm{E}(\bm{k}_3)\rangle&= (2\mathrm{\pi})^3\delta_\mathrm{D}^3(\bm{k}_1+\bm{k}_2+\bm{k}_3)B_{\delta\delta\mathrm{E}}(\bm{k}_1,\bm{k}_2,\bm{k}_3)\label{eq:BSDDE}\\
\langle\tilde{\delta}(\bm{k}_1)\tilde{\gamma}_\mathrm{E}(\bm{k}_2)\tilde{\gamma}_\mathrm{E}(\bm{k}_3)\rangle&= (2\mathrm{\pi})^3\delta_\mathrm{D}^3(\bm{k}_1+\bm{k}_2+\bm{k}_3)B_{\delta\mathrm{E}\mathrm{E}}(\bm{k}_1,\bm{k}_2,\bm{k}_3)\label{eq:BSDEE}\\
\langle\tilde{\gamma}_\mathrm{E}(\bm{k}_1)\tilde{\gamma}_\mathrm{E}(\bm{k}_2)\tilde{\gamma}_\mathrm{E}(\bm{k}_3)\rangle
&= (2\mathrm{\pi})^3\delta_\mathrm{D}^3(\bm{k}_1+\bm{k}_2+\bm{k}_3)B_{\mathrm{E}\mathrm{E}\mathrm{E}}(\bm{k}_1,\bm{k}_2,\bm{k}_3)\label{eq:BSEEE}\\
\langle\tilde{\delta}(\bm{k}_1)\tilde{\gamma}_\mathrm{B}(\bm{k}_2)\tilde{\gamma}_\mathrm{B}(\bm{k}_3)\rangle&= (2\mathrm{\pi})^3\delta_\mathrm{D}^3(\bm{k}_1+\bm{k}_2+\bm{k}_3)B_{\delta\mathrm{B}\mathrm{B}}(\bm{k}_1,\bm{k}_2,\bm{k}_3)\label{eq:BSDBB}\\
\langle\tilde{\gamma}_\mathrm{E}(\bm{k}_1)\tilde{\gamma}_\mathrm{B}(\bm{k}_2)\tilde{\gamma}_\mathrm{B}(\bm{k}_3)\rangle
&= (2\mathrm{\pi})^3\delta_\mathrm{D}^3(\bm{k}_1+\bm{k}_2+\bm{k}_3)B_{\mathrm{E}\mathrm{B}\mathrm{B}}(\bm{k}_1,\bm{k}_2,\bm{k}_3)\ .\label{eq:BSEBB}
\end{align}

\subsection{Methodology for measuring power spectra and bispectra from simulations }\label{sect:sim_meas}

To measure the power spectra and bispectra we follow the methodology in \citet{kurita2021power}. We refer the reader to \citet{kurita2021power} and \citet{shi2021power} for more detailed descriptions of the  measurement methodology. To measure power spectra we use the publicly available package \textsc{nbodykit}\footnote{https://github.com/bccp/nbodykit} \citep{hand2018nbodykit} which provides a wide range of tools to analyse cosmological simulations.  For bispectra we adapt  \textsc{bskit}\footnote{https://github.com/sjforeman/bskit}, developed by   \citet{foreman2020baryonic}.  This package is based on  \textsc{nbodykit} together with the Fast Fourier Transform-based bispectrum measurement algorithm presented  in \citet{tomlinson2019efficient}.   In both cases we incorporate equations~(\ref{eq:gammaE}) and (\ref{eq:gammaB}) into the existing code in order to measure the IA spectra.  

In measuring from simulations we align all axes in Fourier space with the simulation box sides.  We then define a regular grid of size 512$^3$ within the box and assign  the subhalo shape measurements to the grid using the cloud-in-cell interpolation algorithm. 
We investigated the alternative triangle shaped cloud assignment method but found it made little difference to the results.  
We confine our bispectrum measurements to equilateral triangles and to a representative isosceles configuration whose sides have magnitudes in the ratio 2:2:1.   
To quantify uncertainty in the estimates, we divide the simulation box into $3^3$ subboxes, and estimate standard errors using jackknife sampling, excluding each subbox in turn.  

In all subsequent results we show IA auto power spectra with Poisson shot noise subtracted, but without any allowance for non-linear effects due to intrinsic alignments of shapes.   \citet{kurita2021power} explored the latter in detail but found it to be only 5-10 per cent of Poisson shot noise in their halo sample from the Dark Quest simulations.  Thus  we measure shot noise in the IA power spectra as $\epsilon_\mathrm{rms}^2/ n_\mathrm{eff}$ where \mbox{$n_\mathrm{eff}=n_\mathrm{h}/ L_\mathrm{box}^3$} is the effective number density within the simulation box with side length \mbox{$ L_\mathrm{box}$} and $n_\mathrm{h}$ is the number of subhalos in the box. 

 The measured shear \lq fields\rq \ are weighted by the number density of halos. Density weighting is important because halos are biased tracers of the matter density field and ellipticity/shear measurements can only be made at the positions where halos exist.

\section{Analytical modelling}\label{sect:LAM}
Two-point IA statistics are commonly modelled by the linear alignment model \citep{hirata2004intrinsic}.  This model  
assumes that the intrinsic ellipticity of a halo is linearly related to the local quadrupole of the gravitational potential at the redshift at which the halo formed.  Thus in Fourier space we can write 
\begin{align}
\tilde{\gamma}_{({+/\times})}(\bm{k},z) &= -A_{\mathrm{IA}}f_{({+/\times})}\frac{C_1\Omega_\mathrm{m}\rho_\mathrm{cr}}{D(z)} \tilde{\delta}(\bm{k},z)\ ,\label{eq:LA}
\end{align}
where $\Omega_\mathrm{m}$ is the total matter density parameter, $\rho_\mathrm{cr}$ is the critical density at the present day and $D(z)$ is the linear growth factor.  The functions $f_{(+/\times)}$ are defined as 
\begin{align}
f_+&= (1-\mu^2) \cos 2\phi \ ,\\
f_\times &= (1-\mu^2) \sin 2\phi\ ,
\end{align}
where, as before, $\cos^{-1} (\mu)$ is the angle between $\bm{k}$ and the \mbox{$k_z$-axis}  and $\phi = \tan^{-1}(k_x/k_y)$.   The parameter $C_1$ in equation (\ref{eq:LA}) is a normalization factor  which in principle can be determined from observations or simulations. The amplitude $A_\mathrm{IA}$ quantifies the magnitude of the intrinsic alignment effect.  This quantity, commonly used in cosmological inference, is what we are particularly interested in. 

Substituting from equation (\ref{eq:LA}) into equations~(\ref{eq:gammaE}) and  (\ref{eq:gammaB}) we have
\begin{align}
\tilde{\gamma}_\mathrm{E} (\bm{k})&=f_\mathrm{IA}\tilde{\delta}(\bm{k})\ ,\label{eq:gammaE_fIA}\\
\tilde{\gamma}_\mathrm{B} (\bm{k})&= 0\ ,
\end{align}
where we define $f_\mathrm{IA}$ as
\begin{align}
f_\mathrm{IA}&=-A_{\mathrm{IA}}(1-\mu^2)\frac{C_1\Omega_\mathrm{m}\rho_\mathrm{cr}}{D(z)} \ .\label{eq:fIA}
\end{align}

From equation (\ref{eq:gammaE_fIA}) we can obtain 
the three-dimensional E-mode intrinsic alignment power spectra 
\begin{align}
P_{\delta\mathrm{E}}(k)&= f_{\mathrm{IA}}P_\mathrm{NL}(k)\ ,\label{eq:PSDELA}\\
P_{\mathrm{E}\mathrm{E}}(k)&=f_{\mathrm{IA}}^2P_\mathrm{NL}(k)\ ,\label{eq:PSEELA}
\end{align}
 where we have used the non-linear matter power spectrum, $P_\mathrm{NL}(k)$, as suggested by \citet{bridle2007dark}. This modification, known as the non-linear alignment model, has been found empirically to improve the fit of the model at non-linear scales.
 
Theoretically the model predicts that B-mode power spectra are zero to first order, although as discussed previously, $P_\mathrm{BB}$ may be non-zero in practice due to shape noise or to higher-order non-linear contributions, although these would be small \citep{blazek2019beyond}.
 
Note that equations~(\ref{eq:PSDELA}) and (\ref{eq:PSEELA})  also depend on redshift but for brevity we have omitted the $z$-dependence, here and in all subsequent related equations.

\subsection{Extension of linear alignment model to bispectra}
To develop a similar analytical model for three-point statistics we need to relate the intrinsic alignment bispectra to the non-linear matter bispectrum.  It is most straightforward to use a fitting function for the matter bispectrum based on tree-level perturbation theory, for example those given in \citet{scoccimarro2001fitting} and
\citet{gil2012improved}. These have the generic form
\begin{align}
B_{\delta\delta \delta}(\bm{k}_1,\bm{k}_2,\bm{k}_3)&= 2\, [ F_2^{\mathrm{eff}}(\bm{k}_1,\bm{k}_2) P_\mathrm{NL}(k_1)P_\mathrm{NL}(k_2) + 2  \ \text{perms.}]\ ,
\end{align} 
where $ F_2^{\mathrm{eff}}(\bm{k}_1,\bm{k}_2)$ is a modification of the standard perturbation theory kernel \citep{bernardeau2002large}.

This formulation is easily extended to include  IA power spectra in place of the non-linear matter power spectrum, leading directly to expressions for the intrinsic alignment bispectra which are in the spirit of the two-point non-linear alignment model \citep{pyne2021self}:
\begin{align}
B_{\delta\delta  \mathrm{E}}(\bm{k}_1,\bm{k}_2,\bm{k}_3)&=
 2\, \Big[ f_{\mathrm{IA}}^2F_2^{\mathrm{eff}}(\bm{k}_1,\bm{k}_2) P_\mathrm{NL}(k_1)P_\mathrm{NL}(k_2)\label{eq:BSDDELA}\\ 
 &\hspace{0.4cm}+ f_{\mathrm{IA}}F_2^{\mathrm{eff}}(\bm{k}_2,\bm{k}_3) P_\mathrm{NL}(k_2)P_\mathrm{NL}(k_3)\notag\\
&\hspace{0.4cm} +f_{\mathrm{IA}}F_2^{\mathrm{eff}}(\bm{k}_3,\bm{k}_1) P_\mathrm{NL}(k_3)P_\mathrm{NL}(k_1)\Big]\ ,\notag\\
 B_{\delta  \mathrm{E} \mathrm{E}}(\bm{k}_1,\bm{k}_2,\bm{k}_3)
 &=2\, \Big[ f_{\mathrm{IA}}^3F_2^{\mathrm{eff}}(\bm{k}_1,\bm{k}_2) P_\mathrm{NL}(k_1)P_\mathrm{NL}(k_2)\label{eq:BSDEELA}\\
 &\hspace{0.4cm}+ f_{\mathrm{IA}}^2F_2^{\mathrm{eff}}(\bm{k}_2,\bm{k}_3) P_\mathrm{NL}(k_2)P_\mathrm{NL}(k_3)\notag\\
 &\hspace{0.4cm}+f_{\mathrm{IA}}^3F_2^{\mathrm{eff}}(\bm{k}_3,\bm{k}_1) P_\mathrm{NL}(k_3)P_\mathrm{NL}(k_1)\Big] \ , \notag\\
  B_{ \mathrm{E} \mathrm{E}\mathrm{E}}(\bm{k}_1,\bm{k}_2,\bm{k}_3)&= f_{\mathrm{IA}}^4 B_{\delta\delta\delta}(\bm{k}_1,\bm{k}_2,\bm{k}_3)\ .\label{eq:BSEEELA}
 \end{align}
 The bispectrum $B_{ \mathrm{E} \mathrm{E}\mathrm{E}}$ is positive but  the signs of $B_{\delta\delta  \mathrm{E}}$ and $B_{\delta  \mathrm{E} \mathrm{E}}$ depend on the triangle configuration.
 
 For equilateral triangles equations~(\ref{eq:BSDDELA}) to (\ref{eq:BSEEELA})  reduce to
\begin{align}
B_{\delta\delta \mathrm{E}} &= \frac{1}{3}\Big[f_{\mathrm{IA}}^2 +2 f_{\mathrm{IA}}\Big] B_{\delta\delta\delta}\ ,\label{eq:BSequiDDE}\\
B_{\delta  \mathrm{E} \mathrm{E}}&=\frac{1}{3}\Big[2f_{\mathrm{IA}}^3+f_{\mathrm{IA}}^2\Big] B_{\delta\delta\delta}\label{eq:BSequiDEE}\ ,\\
B_{ \mathrm{E} \mathrm{E} \mathrm{E}}&=  f_{\mathrm{IA}}^4B_{\delta\delta\delta}\ ,\label{eq:BSequiEEE}
\end{align}
where we have omitted the $\bm{k}$ arguments for brevity.

In this work we use the fitting formula from \citet{gil2012improved} in equations~(\ref{eq:BSDDELA}) to (\ref{eq:BSEEELA}) with the non-linear matter power spectrum estimated by the fitting formula in \citet{takahashi2012revising}. We note, however, 
 that the formulas of both \citet{scoccimarro2001fitting} and 
\citet{gil2012improved} are known to have deficiencies: they are fitted over  limited $k$ scales and have been found to be inaccurate for squeezed triangles \citep{namikawa2019cmb}. These issues were explored in detail by \citet{takahashi2020fitting} who developed a new formula, Bihalofit, based on the halo model. It is more accurate than the perturbation theory based fitting functions over a wider range of wavenumbers and redshifts.  However it does not lend itself to use in our model because it is the sum of 1-halo and 3-halo terms which involve halo model integrals and a large number of fitted parameters. Thus it cannot easily be related to the NLA model.  However, as we show in Appendix \ref{sect:BHF_comp}, for scales up to at least $k \approx 3 \, h \mathrm{Mpc}^{-1}$ there is good agreement between the matter bispectrum estimated from \citet{gil2012improved} and from \citet{takahashi2020fitting} and between these results and our measurements from simulations.

\section{Results}\label{sect:results}
We present our results as follows. First, in Section~\ref{sect:ell_mass}, we discuss the  halo ellipticity distributions which underlie our intrinsic alignment measurements.  In Section~\ref{sect:spectra} we present our measurements of intrinsic alignment power spectra and bispectra from simulations, and discuss consistency with previous power spectrum results and between our two-point and three-point measurements. In the rest of Section~\ref{sect:results} we use these simulation measurements to validate our  analytical models.  In Sections~\ref{sect:IAPS} and \ref{sect:IAmass} we focus on IA power spectra, using the NLA model to estimate intrinsic alignment amplitudes and explore their mass dependence. These power spectrum results establish our methodology and provide context for our principal results from bispectra based on our three-point analytical IA model.  Results for equilateral triangles are given in  Section~\ref{sect:IABS} and compared to power spectrum results in Section~\ref{sect:compPSBS}. Finally, in  Section~\ref{sect:isos_BS} we discuss non-equilateral IA bispectra.

\subsection{Distribution of halo ellipticities by mass}\label{sect:ell_mass}

Previous measurements of halo intrinsic alignments from simulations, for example \citet{jing2002intrinsic}, \citet{lee2008quantifying}, \citet{xia2017halo}, \citet{piras2018mass} and \citet{kurita2021power}, have found that intrinsic alignment increases with increasing halo mass and to a lesser extent with redshift. Similar trends have  been noted from  measurements of galaxy shapes from simulations \citep{,tenneti2014galaxy,tenneti2015intrinsic}, and in survey data for galaxies \citep{joachimi2011constraints,joachimi2013intrinsic,singh2015intrinsic} and clusters \citep{van2017intrinsic}.  In view of this  we split our measured subhalos into four mass bins, with each bin spanning one order of magnitude from $10^{10}$ to $10^{14} \ h^{-1}\mathrm{M}_\odot$.  

\begin{figure}
\includegraphics[width=8cm]{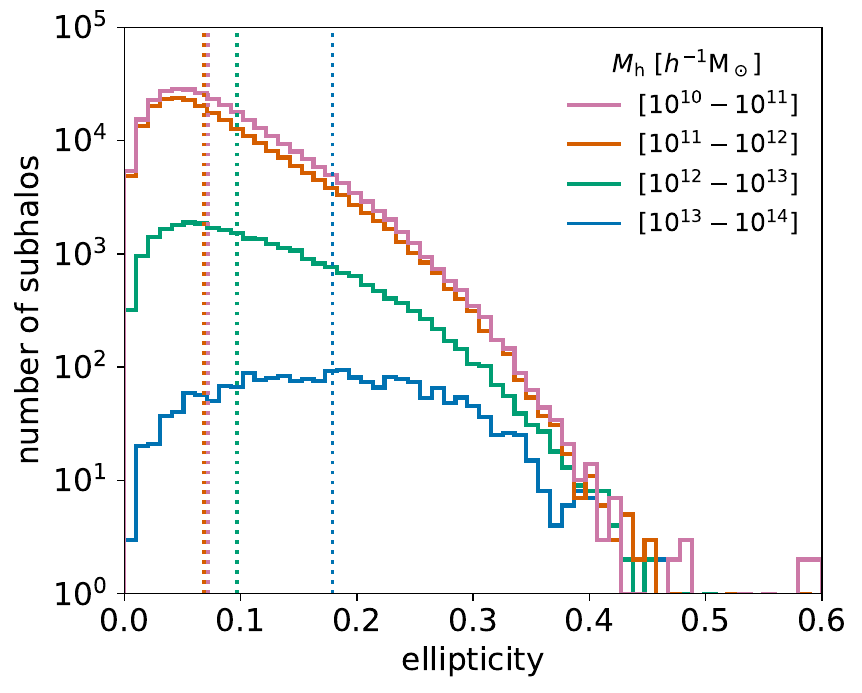} 
\caption{Distribution of dark matter subhalos by projected ellipticity \mbox{$\epsilon = \sqrt{\epsilon_+^2+\epsilon_\times^2}$} in four mass bins at $z=0.5$. Vertical dotted lines indicate the median ellipticity in each bin. }\label{fig:ellipticity} 
\end{figure}

Figure~\ref{fig:ellipticity}  shows the distribution of ellipticity, defined as \mbox{$\epsilon = \sqrt{\epsilon_+^2+\epsilon_\times^2}$}, in each of the four bins at $z=0.5$. (Similar distributions are found for other redshifts.)   Dotted vertical lines show the median ellipticity in each bin.   
This confirms previous authors' findings: higher-mass halos are more elliptical.  
It is noticeable that the two lowest-mass bins have similar ellipticities, which are lower than those of the two high-mass bins.

\subsection{IA power spectra and bispectra measured from simulations}\label{sect:spectra}

\begin{figure*}
\centering

\includegraphics[width=15cm]{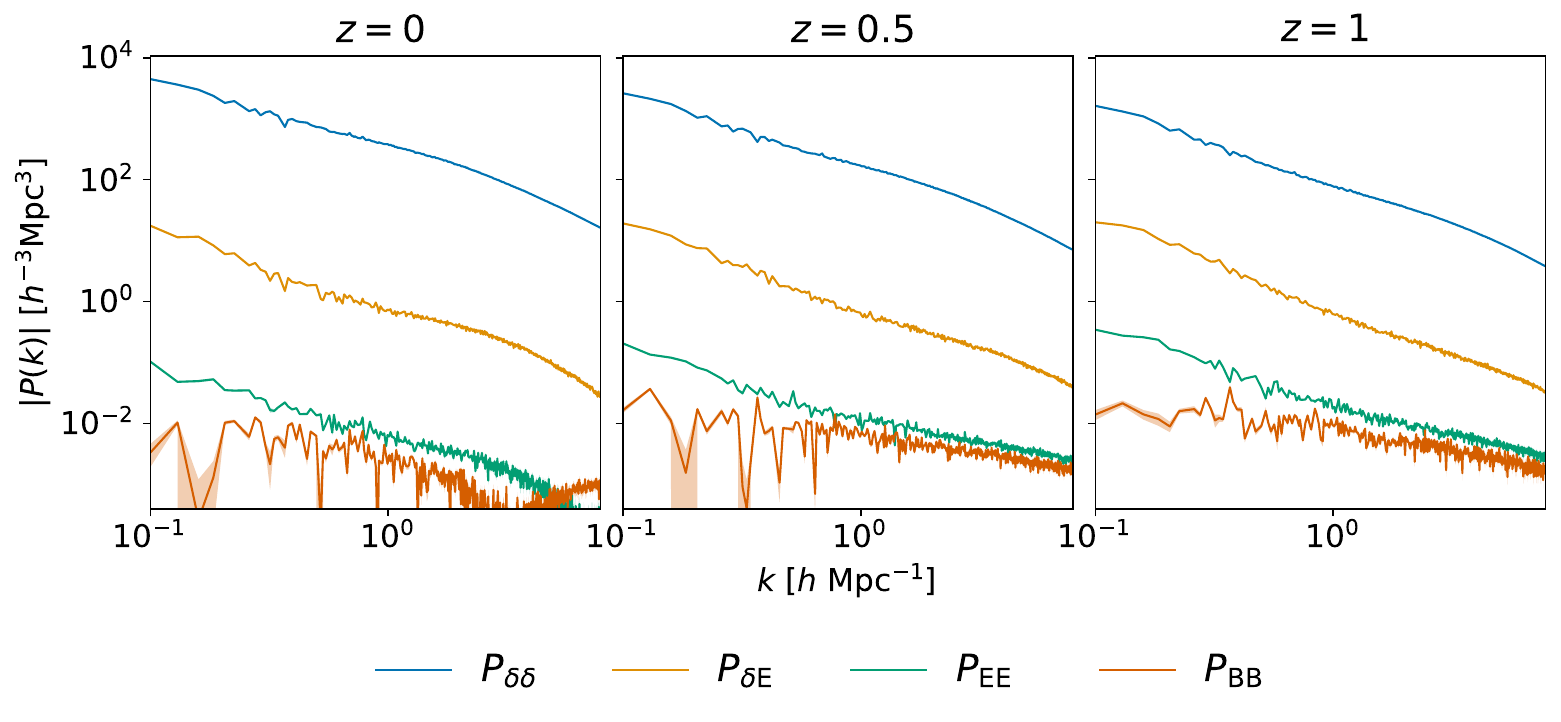}
\caption{  Intrinsic alignment power spectra $P_{\delta\mathrm{E}}, P_{\mathrm{E}\mathrm{E}}$ and $P_{\mathrm{B}\mathrm{B}}$ measured from simulations.  The matter power spectrum $P_{\delta\delta}$ is also shown for reference. Shot noise has been subtracted from the auto power spectra. Shaded areas are 68\% confidence intervals. }\label{fig:IAPS_errors} 
\end{figure*}

Figure \ref{fig:IAPS_errors} shows our measured intrinsic alignment power spectra, $P_{\delta\mathrm{E}}$, $P_{\mathrm{E}\mathrm{E}}$  and $P_{\mathrm{B}\mathrm{B}}$, defined by equations~(\ref{eq:PSDE}) to (\ref{eq:PSBB}),   at three redshifts, with the non-linear matter power spectrum $P_{\delta\delta}$ shown for comparison. In this and all similar figures we show the absolute value of the spectra. 
$P_{\mathrm{B}\mathrm{B}}$ is essentially equal to the shot noise at all but the largest scales considered here.   The non-vanishing cross-power spectrum $P_{\delta\mathrm{E}}$ confirms that  subhalo shapes are correlated with the matter overdensity field at all scales.  The power spectra show little variation across redshifts.  These results are consistent with IA power spectrum measurements in \citet{kurita2021power}. (See their figs 2 and 4).

\begin{figure*}
\centering

\includegraphics[width=15cm]{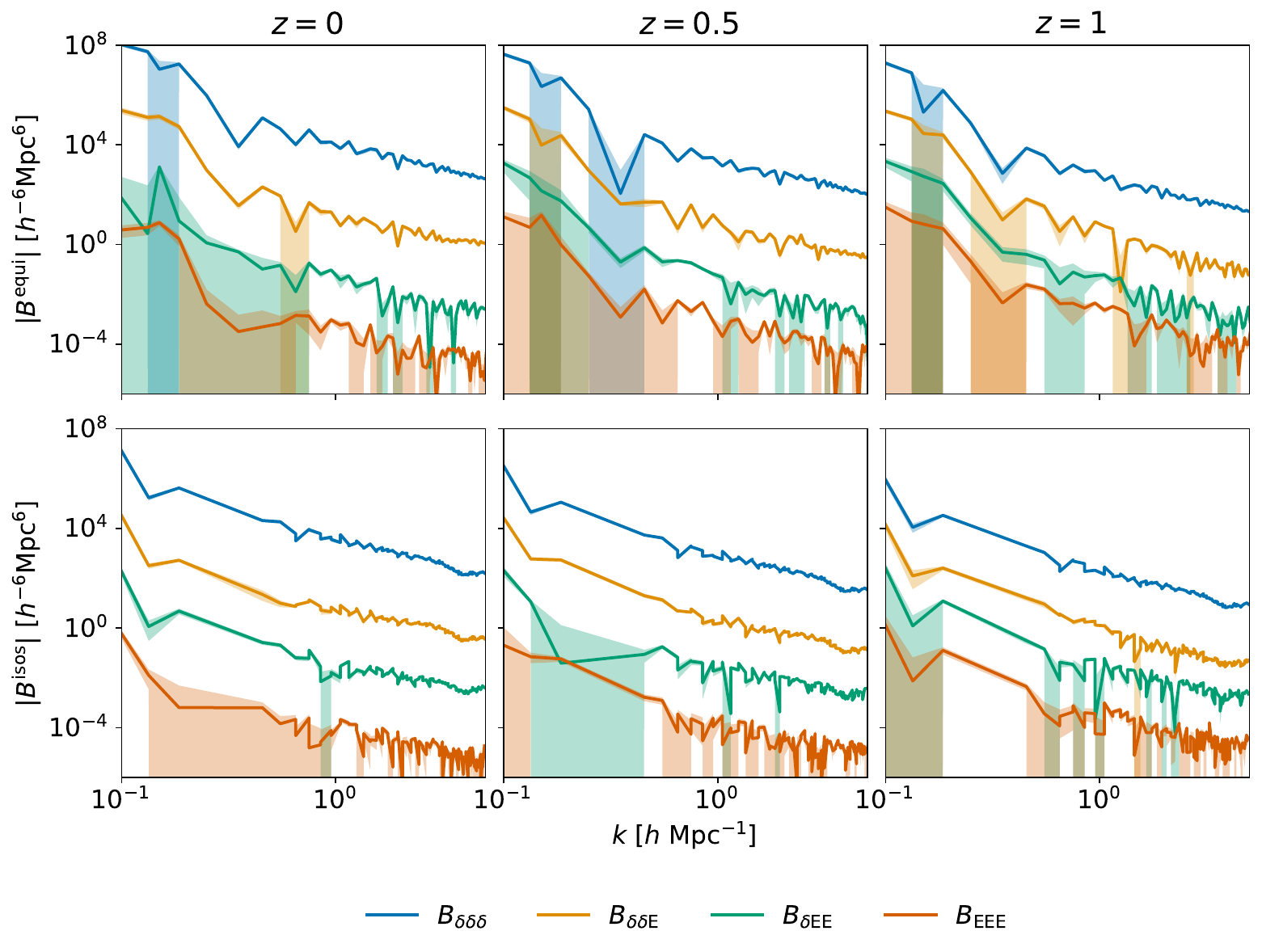}

\caption{ Intrinsic alignment  bispectra $B_{\delta\delta\mathrm{E}}, B_{\delta\mathrm{E}\mathrm{E}}$ and $B_{\mathrm{E}\mathrm{E}\mathrm{E}}$ measured from simulations. \textit{Top}: Equilateral triangles.  \textit{Bottom}: Isosceles triangles with sides in the ratio 2:2:1.  The  bispectrum $B_{\delta\delta\delta}$ is also shown for reference. Shaded areas are 68\% confidence intervals. }\label{fig:IABS_errors} 
\end{figure*}

\begin{figure*}
\centering

\includegraphics[width=15cm]{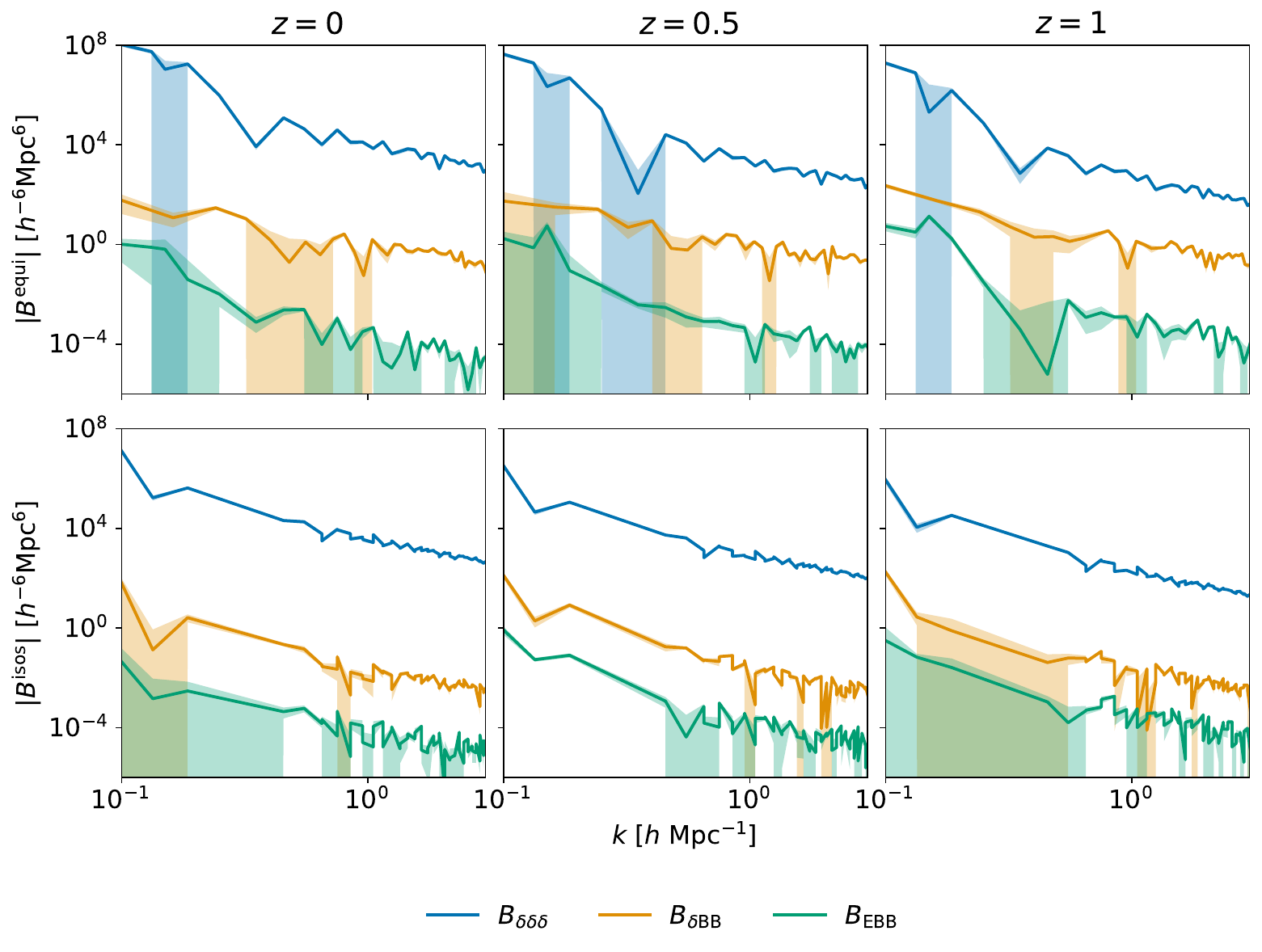}

\caption{ Intrinsic alignment bispectra $B_{\delta\delta\mathrm{B}}$ and $B_{\mathrm{E}\mathrm{B}\mathrm{B}}$ measured from simulations. \textit{Top}: Equilateral triangles.  \textit{Bottom}: Isosceles triangles with sides in the ratio 2:2:1.  The bispectrum $B_{\delta\delta\delta}$ is also shown for reference. Shaded areas are  68\% confidence intervals.  }\label{fig:IABS_errors_B} 
\end{figure*}

Figures~\ref{fig:IABS_errors} and  \ref{fig:IABS_errors_B} show similar results for measured E-mode and B-mode bispectra respectively (equations~(\ref{eq:BSDDE}) to (\ref{eq:BSEBB})), for equilateral triangles and for isosceles triangles with sides in the ratio 2:2:1. The B-mode bispectra have very low signal-to-noise ratios (see Appendix \ref{sect:StoN}) and we do not consider them further in this work. However the E-mode bispectra, in particular  $B_{\delta\delta\mathrm{E}}$, have relatively strong signals, especially for isosceles configurations.  
 
There is no convenient benchmark with which to compare our bispectrum results. The only previous measurements of three-point intrinsic alignments from simulations were by \citet{semboloni2008sources} and \citet{semboloni2010weak}. Unfortunately it is difficult to compare our results with theirs for a number of reasons: they worked in configuration space and measured aperture mass statistics, and they focused on the  magnitudes of intrinsic alignments relative to lensing signals, rather than on the strength  of the IA signal itself.  This is discussed further in Section \ref{sect:comp}.

\begin{figure*}
\centering

\begin{subfigure}[b]{1.\linewidth}
\hspace{1.2cm}
\includegraphics[width=15cm]{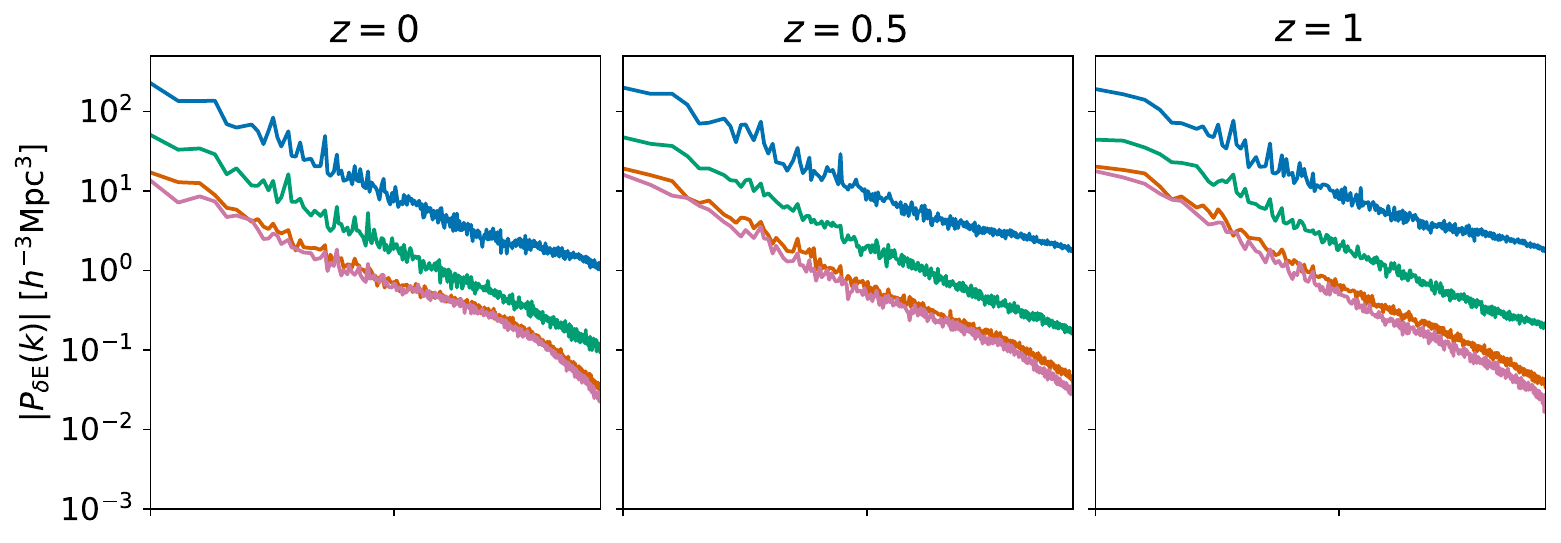}
\end{subfigure}
\begin{subfigure}[b]{1.\linewidth}
\hspace{1.2cm}

\hspace{1.2cm}
\includegraphics[width=15cm]{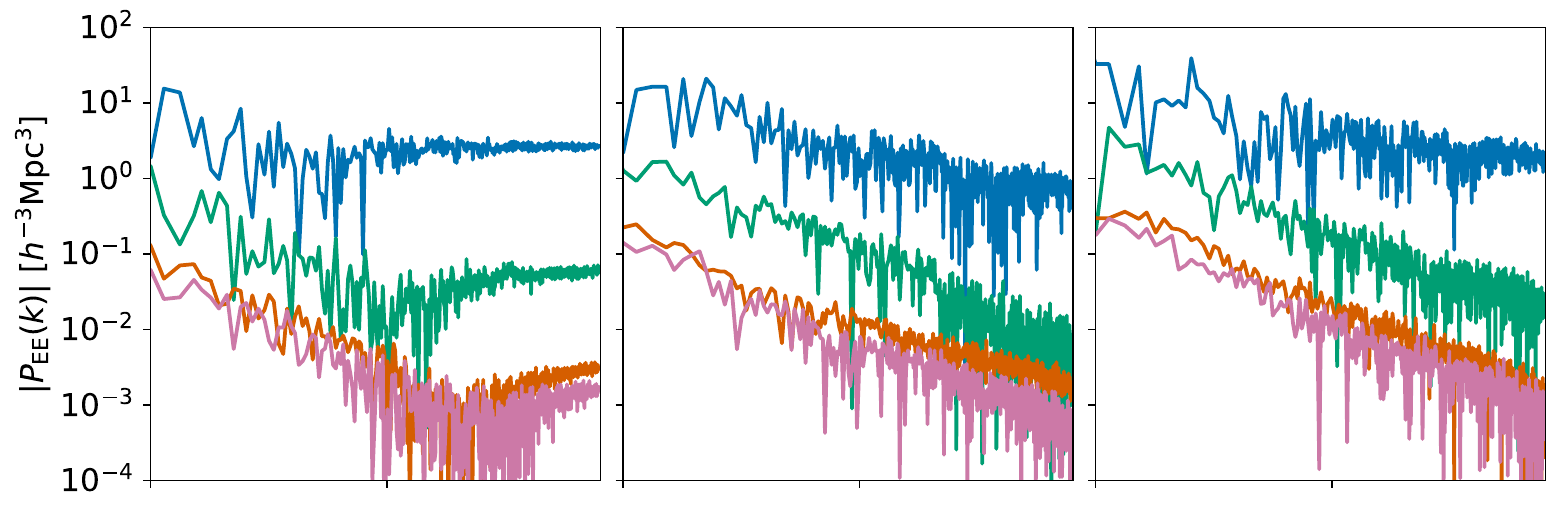}
\end{subfigure}

\begin{subfigure}[b]{1.\linewidth}
\hspace{1.2cm}
\includegraphics[width=15cm]{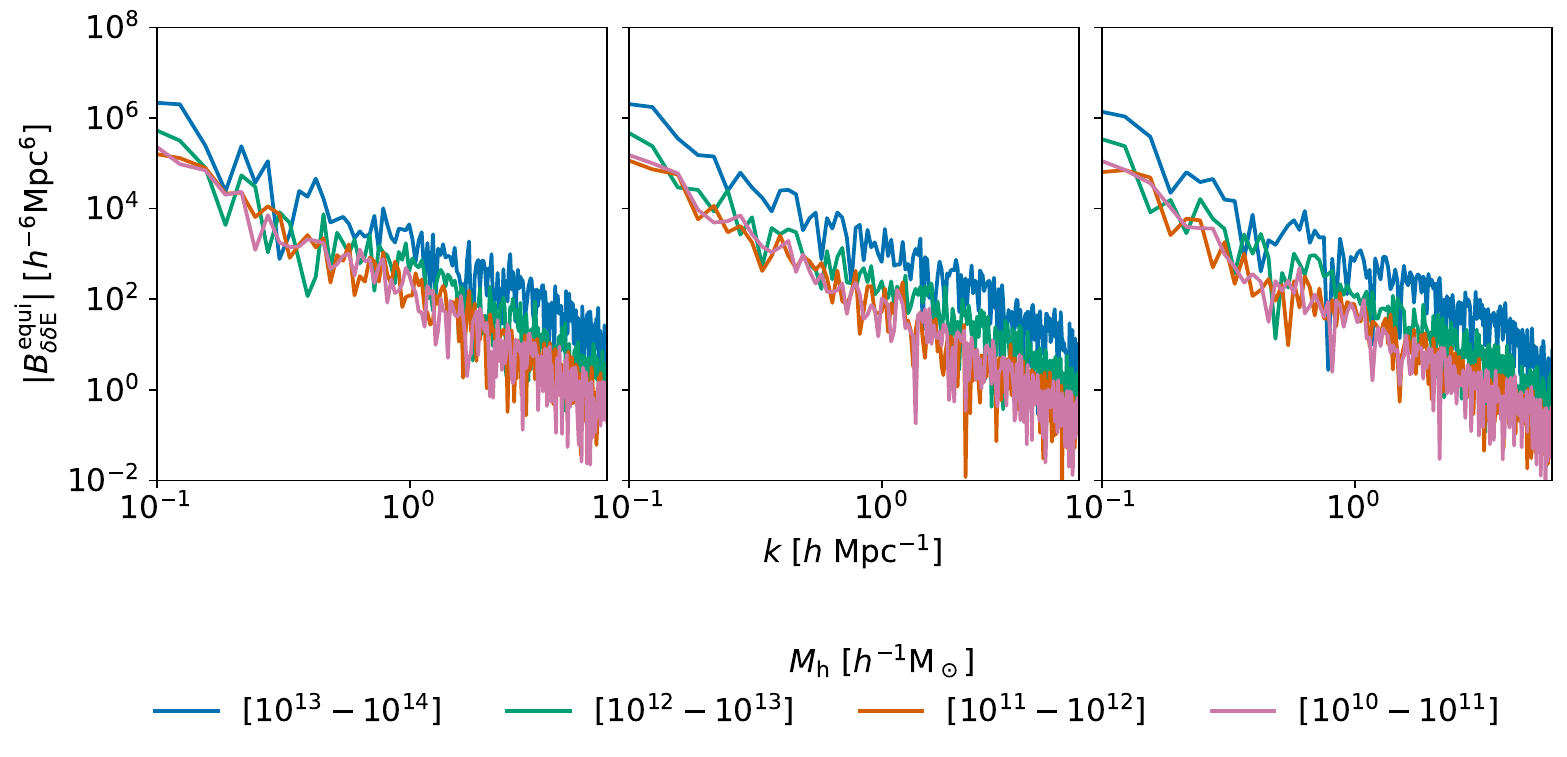}
 \end{subfigure}  

 \caption{    Intrinsic alignment spectra in four mass bins. \textit{Top}: $P_{\delta \mathrm{E}}$, \textit{Centre}: $P_{\mathrm{EE}}$, \textit{Bottom}:  $B^\mathrm{equi}_{\delta\delta\mathrm{E}}$.  }\label{fig:IAPSBS_mass} 

\end{figure*}

In  Fig.~\ref{fig:IAPSBS_mass} we split the IA power spectra $P_{\delta \mathrm{E}}$ and $P_{\mathrm{EE}}$  and the bispectrum $B_{\delta\delta\mathrm{E}}$ between mass bins.   In all three  cases the   magnitudes of the spectra increase with mass.  
Again we note that the spectra in the two lowest-mass bins are similar to each other and generally smaller than those in higher-mass bins.

\subsection{Intrinsic alignment amplitude from power spectra}\label{sect:IAPS}
 
Having shown that E-mode spectra are detectable from the simulations, we now estimate the intrinsic alignment amplitude $A_{\mathrm{IA}}$ in equation (\ref{eq:LA}) from simulation measurements.  We first consider power spectra. From equations~  (\ref{eq:PSDELA}) and (\ref{eq:PSEELA}) we can obtain approximate estimates of  $f_{\mathrm{IA}}$, and hence $A_{\mathrm{IA}}$, from the ratio $P_{\delta\mathrm{E}}(k)/P_{\delta\delta}(k)$ or from $\sqrt{P_{\mathrm{E}\mathrm{E}}(k)/P_{\delta\delta}(k)}$.  
Alternatively, following \citet{kurita2021power}, we can use least-squares minimization to fit  $A_{\mathrm{IA}}$  from these equations.  To do this we find the value of the parameter $\hat{A}_\mathrm{IA}$ which minimizes $\chi^2$ given by
\begin{align}
\chi^2&=\mathlarger{\sum}_k \  {\frac{[R(k)-F(\hat{A}_\mathrm{IA})]^2}{\sigma^2_R(k)}}\ .\label{eq:PSchi2}
\end{align}
Here, in the first case, based on $P_{\delta \mathrm{E}}$,  \mbox{$R(k) = P_{\delta \mathrm{E}}(k)/P_{\delta\delta}(k)$} and 
\begin{align}
F(\hat{A}_\mathrm{IA}) &=\int_0^1 \, (1-\mu^2) \ \mathrm{d}\mu \, c(z)\hat{A}_\mathrm{IA} 
= \frac{2}{3} c(z)\hat{A}_\mathrm{IA} \ , 
\end{align}
where $c(z) = C_1\Omega_\mathrm{m}\rho_\mathrm{cr}/D(z)$ (see equation~(\ref{eq:fIA})) and $\sigma^2_R(k)$ is the variance of $R(k)$ calculated using jackknife sampling from 27 simulation subboxes. To calculate $F(\hat{A}_\mathrm{IA})$ we obtain the growth factor  from  the cosmological parameter estimation code CosmoSIS\footnote{https://bitbucket.org/joezuntz/cosmosis/wiki/Home} \citep{zuntz2015cosmosis}.  We use the value of $C_1$ derived by \citet{bridle2007dark} which is $5\times 10^{-14} \,  h^{-2} \mathrm{M}_{\odot}^{-1}\mathrm{Mpc}^3$, leading to \mbox{$C_1\rho_\mathrm{cr} = 0.0134$} \citep{joachimi2011constraints}.    All cosmological parameters values are identical to those used in the simulations.

In the second case,  based on $P_{\mathrm{EE}}$,  \mbox{$R(k) = \sqrt{P_{\mathrm{EE} }(k)/P_{\delta\delta}(k)}$} and
\begin{align}
F(\hat{A}_\mathrm{IA}) &=\int_0^1\, (1-\mu^2)^2 \ \mathrm{d}\mu \, c(z)\hat{A}_\mathrm{IA} 
= \frac{8}{15}  c(z)\hat{A}_\mathrm{IA} \ , 
\end{align}
with $c(z)$ and  $\sigma^2_R(k)$ defined as before.

\begin{figure*}

\centering

\includegraphics[width=15cm]{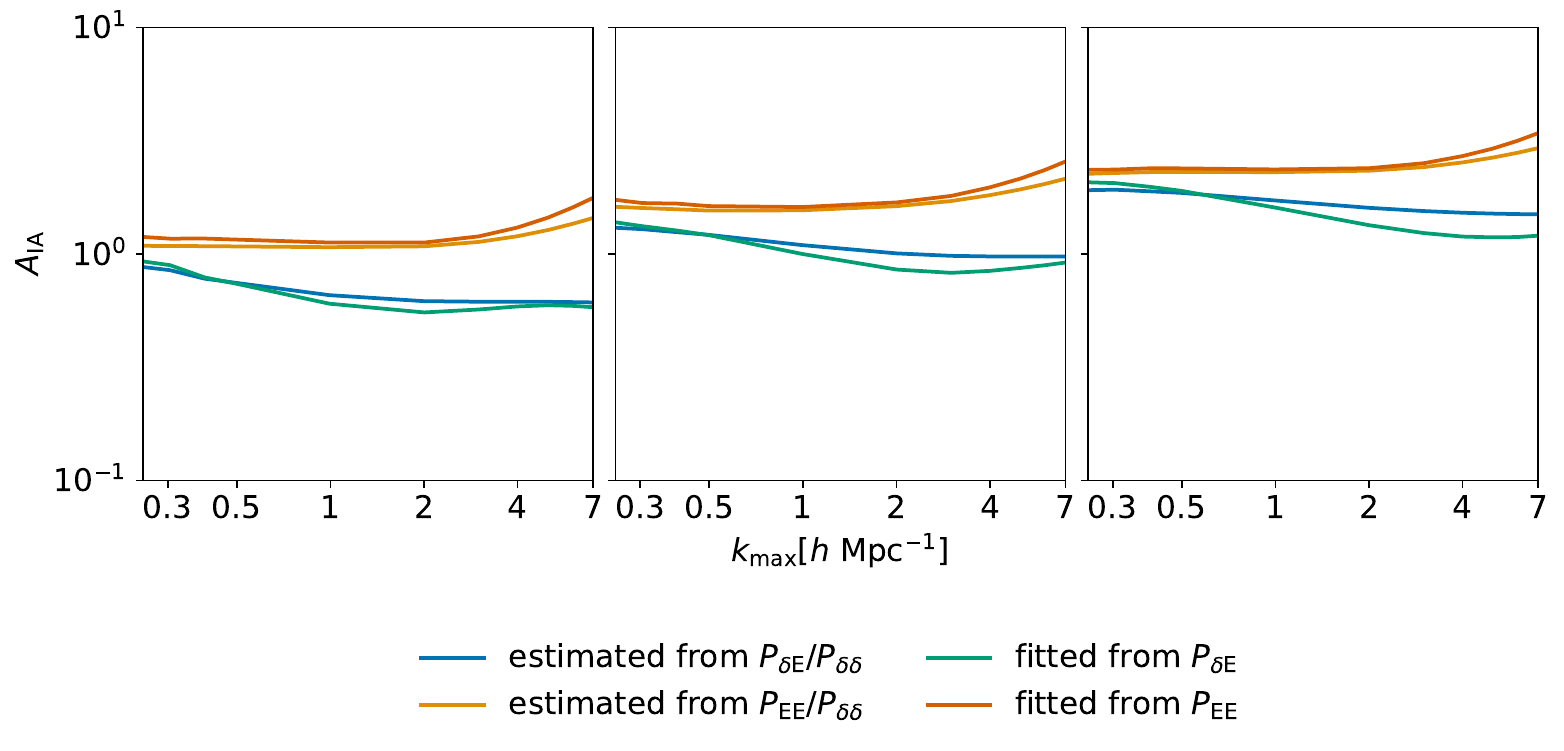}
\caption{ The mean intrinsic alignment amplitude $A_\mathrm{IA}$   as a function of $k_\mathrm{max}$  estimated  from the relationship between the IA power spectra and the matter power spectrum given by equations~(\ref{eq:PSDELA}) and (\ref{eq:PSEELA}) and also fitted using equation (\ref{eq:PSchi2}) over the range $[0.1,k_{\mathrm{max}}] \  h \mathrm{Mpc}^{-1}$. } \label{fig:IAPS_comp} 
\end{figure*}

Figure~\ref{fig:IAPS_comp} compares the values of $A_{\mathrm{IA}}$ obtained from ratios of power spectra to those obtained using equation (\ref{eq:PSchi2}). Results are for \mbox{$k_\mathrm{min} = 0.1 \, h \mathrm{Mpc}^{-1}$} and varying $k_\mathrm{max}$.  There is good agreement for all values of $k_\mathrm{max}$. 
This validates the NLA model. The results for $P_{\mathrm{E}\mathrm{E}}$ are consistently higher than those for $P_{\delta \mathrm{E}}$, suggesting  a possible unexplained systematic trend.

\subsection{Dependence of IA amplitude on mass}\label{sect:IAmass}
Since the intrinsic alignment power spectra depend strongly on subhalo mass, we also expect $A_\mathrm{IA}$ to depend on mass.  We therefore  obtain  estimates of $A_\mathrm{IA}$ from $P_{\delta \mathrm{E}}$ and $P_{\mathrm{E} \mathrm{E}}$ for each of our four mass bins. We assume that the same linear alignment model applies for all halo masses, with any mass-dependence being absorbed into the estimated amplitude.  Alternative ways in which the mass-dependence has been incorporated include inserting a mass-dependent halo bias term into the linear alignment model \citep{xia2017halo} and  using a virial argument to derive a scaling by mass \citep{piras2018mass}.

\begin{figure*}
\centering

 \includegraphics[width=15cm]{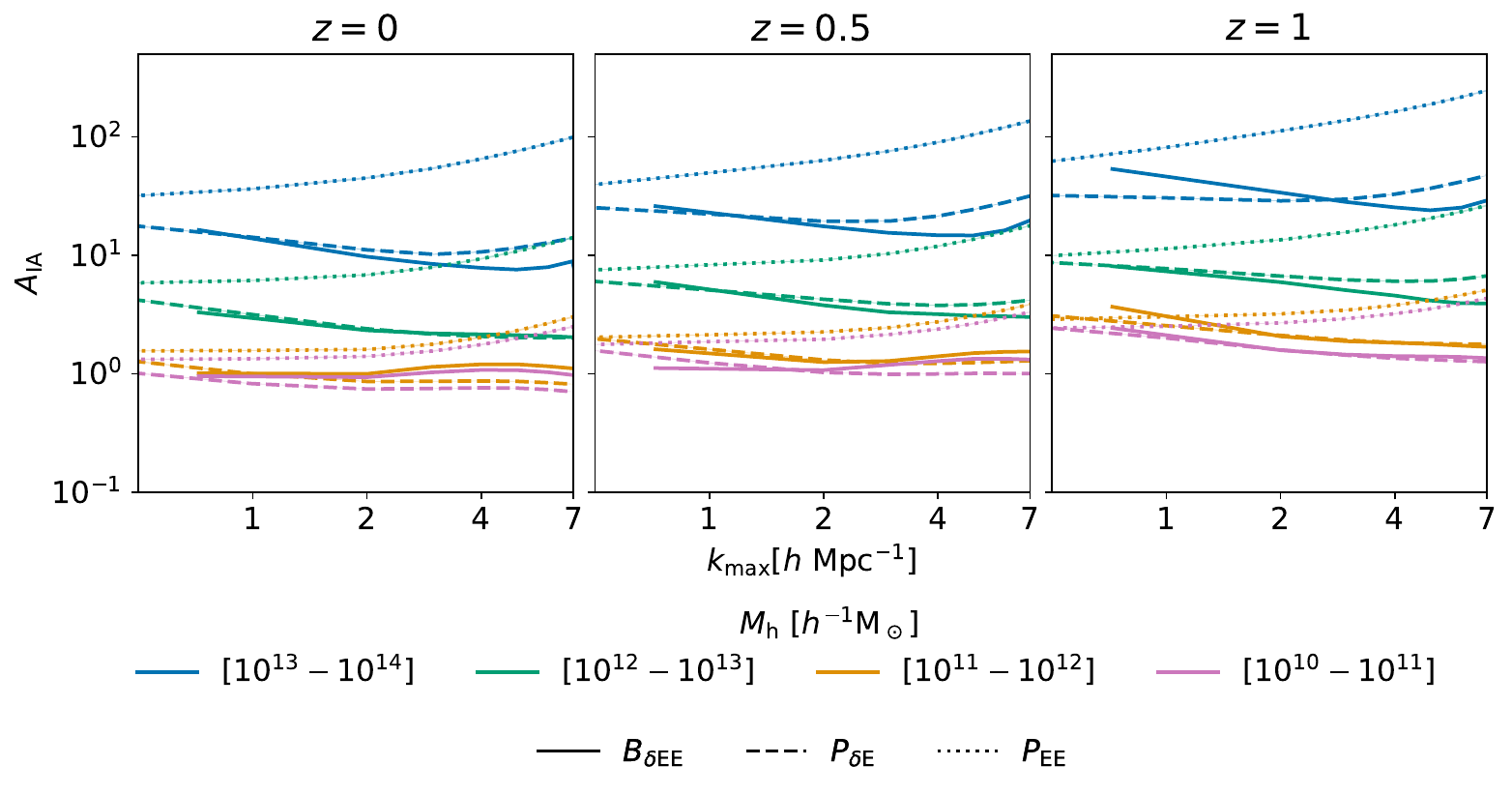}  

\caption{  The intrinsic alignment amplitude $A_\mathrm{IA}$ estimated from $P_{\delta\mathrm{E}}$, $P_{\mathrm{E}\mathrm{E}}$ and $B_{\delta\delta\mathrm{E}}^\mathrm{equi}$ for four subhalo mass ranges, as a function of  $k_\mathrm{max}$. Typical uncertainties in the fits  are given in Table~\ref{tab:A_IA_summary_mass}.}\label{fig:AIAbymass} 
\end{figure*}

We perform the $\chi^2$ minimization  using equation (\ref{eq:PSchi2}) for each of the four mass bins and for three redshifts, and also vary the maximum $k$ value used. The results  are shown in  Fig.~\ref{fig:AIAbymass}. 
The values of $A_\mathrm{IA}$ obtained from $P_{\delta \mathrm{E}}$ are consistent with fig.~6 in \citet{kurita2021power}, given slightly different mass bins and $k$ and $z$ ranges.  

\begin{figure*}
 \includegraphics[width=15cm]{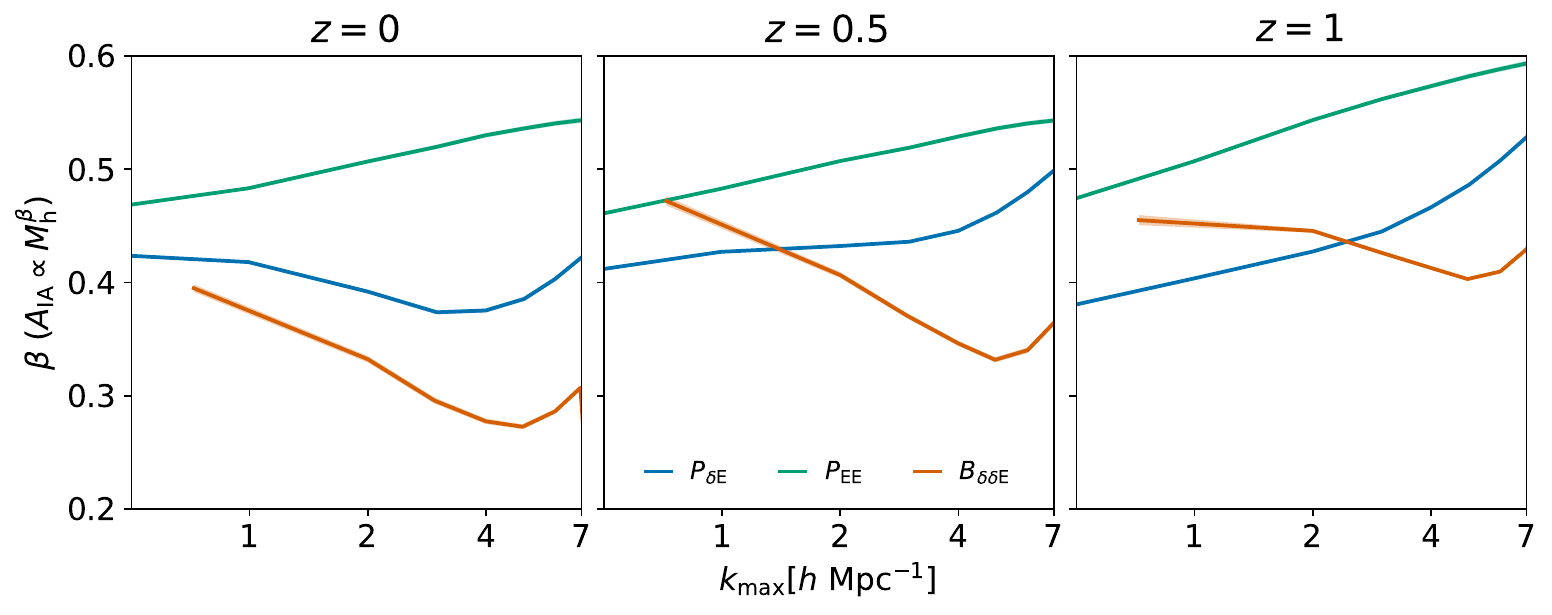}  
\caption{  Slope $\beta$ from a power law relationship $A_\mathrm{IA}\propto M_\mathrm{h}^\beta$,  estimated from $P_{\delta\mathrm{E}}$, $P_{\mathrm{E}\mathrm{E}}$ and $B_{\delta\delta\mathrm{E}}^\mathrm{equi}$. Typical uncertainties in the fits  are given in Table~\ref{tab:A_IA_summary_mass}.}  \label{fig:slopeallz}
\end{figure*}

We also fit a power law of the form 
\begin{align}
A_\mathrm{IA}&\propto M_\mathrm{h}^\beta\ ,\label{eq:beta}
\end{align}
 where $M_\mathrm{h}$ is the mean halo mass per bin.  We estimate this relationship using  both $P_{\delta \mathrm{E}}$ and $P_{\mathrm{E}\mathrm{E}}$.  The results  are shown in  Fig.~\ref{fig:slopeallz}. At all redshifts the two estimates are reasonably consistent for \mbox{$k_\mathrm{max}\le 2\ h \, \mathrm{Mpc}^{-1}$}, but show some variability for higher $k_\mathrm{max}$.  
 
 We can  compare our estimated power spectrum slopes in Fig.~\ref{fig:slopeallz} with those found by \citet{piras2018mass} from the Millennium simulation.  These authors obtained \mbox{$\beta \approx 0.35$} for \mbox{$z=0.46$}, with a slightly lower-mass sample of halos \mbox{($ 10^{11.36}< M_\mathrm{h} <10^{13.36} \ h^{-1}\mathrm{M}_\odot$)}.  Our value \mbox{$\beta \approx 0.43$} for \mbox{$k_\mathrm{max}\approx 2\ h \, \mathrm{Mpc}^{-1}$} at $z=0.5$ is somewhat higher. We note, however, that \citet{piras2018mass} repeated their analysis using observational data and found $\beta \approx 0.56$. They suggested several possible reasons for this difference, including the effect of baryons on halo shapes and the relative strength of the stellar and dark matter signals in different mass bins.  
  
\subsection{Intrinsic alignment amplitude from equilateral bispectra}\label{sect:IABS}
We now turn to our main aim: to investigate whether our three-point analytical model is consistent with the two-point NLA model over the non-linear scales of interest.
We start with the equilateral bispectrum $B_{\delta\delta\mathrm{E}}^\mathrm{equi}$ given by equation (\ref{eq:BSequiDDE}).  We later expand this to non-equilateral versions of $B_{\delta\delta\mathrm{E}}$, but do not consider other IA bispectra because we find these have insufficient signal to provide robust estimates of the  amplitude $A_\mathrm{IA}$.

We again use equation (\ref{eq:PSchi2}) to fit a parameter  $\hat{A}_\mathrm{IA}$ but in this case \mbox{$R(k) = B_{\delta\delta \mathrm{E}}^\mathrm{equi}/B_{\delta\delta\delta}$}. Also, using equation (\ref{eq:BSequiDDE}), we now have 
\begin{align}
F(\hat{A}_\mathrm{IA})&= \int_0^1 \, (1-\mu^2) \ \mathrm{d}\mu \, \frac{ [f_\mathrm{IA}^2 + 2f_\mathrm{IA}]}{3}\\
&= \frac{2}{9} [(c(z)\hat{A}_\mathrm{IA})^2 + 2c(z)\hat{A}_\mathrm{IA}]\ ,
\end{align}
where $c(z)$ and $\hat{A}_\mathrm{IA}$ are defined as in Section~ \ref{sect:IAPS}.
In contrast to the power spectra, there is no simple relationship between the ratio  \mbox{$B_{\delta\delta \mathrm{E}}^\mathrm{equi}/B_{\delta\delta\delta}$} and the intrinsic alignment amplitude so  we cannot produce a plot similar to  Fig.~\ref{fig:IAPS_comp}.

As shown in Fig.~\ref{fig:IAPSBS_mass}, $B_{\delta\delta \mathrm{E}}^\mathrm{equi}$ increases with mass, albeit not as consistently as $P_{\delta\mathrm{E}}$, so we again obtain estimates of $A_\mathrm{IA}$ for each mass bin, and fit a power law to this.
Figures \ref{fig:AIAbymass} and \ref{fig:slopeallz} show the resulting estimates of $A_{\mathrm{IA}}$ and of the slope $\beta$.  
Note that we fit the bispectrum measurements only for  \mbox{$k\gtrsim 0.7  \ h \, \mathrm{Mpc}^{-1}$}  in contrast to  \mbox{$k\gtrsim 0.1  \ h \, \mathrm{Mpc}^{-1}$} for the power spectrum.

\subsection{Comparison of results from power spectra and equilateral bispectra}\label{sect:compPSBS}

Table~\ref{tab:A_IA_summary_mass} summarizes our main results. It shows estimates of $A_\mathrm{IA}$  derived from $P_{\delta\mathrm{E}}$, $P_{\mathrm{E}\mathrm{E}}$ and  $B_{\delta\delta\mathrm{E}}^\mathrm{equi}$ for each mass bin.   Also in this table are the estimated power-law slopes $\beta$ from equation (\ref{eq:beta}) and the mean value of $A_\mathrm{IA}$, weighted by the number of subhalos per mass bin.  For illustrative purposes we choose  \mbox{$k_\mathrm{max}=2 \ h \, \mathrm{Mpc}^{-1}$} in this table, well within non-linear scales,  since our models appear reliable up to this value. This range also covers the typical fit range for cosmic shear studies.   The overall picture would be qualitatively similar with a different  $k_\mathrm{max}$.  

The IA amplitudes obtained from $P_{\delta\mathrm{E}}$ and $B_{\delta\delta\mathrm{E}}$ are completely consistent, underlining the validity of the analytical modelling  up to at least $k_\mathrm{max}=2 \ h \, \mathrm{Mpc}^{-1}$ and confirming that our modelling of three-point IA statistics is consistent with 
the two-point non-linear alignment model. 
The power-law relationship which we obtain using $B_{\delta\delta\mathrm{E}}^\mathrm{equi}$ is broadly consistent with the power spectrum results but is rather lower in most cases and more dependent on the value of $k_\mathrm{max}$.

  \begin{table*}
\vspace{0.9cm}
\caption{Estimates of the intrinsic alignment amplitude $A_\mathrm{IA}$ and the power-law exponent $\beta$ in equation (\ref{eq:beta})  for subhalos in four mass ranges, with 68 per cent confidence intervals, from  the power spectra $P_{\delta \mathrm{E}}$ and  $P_{ \mathrm{E} \mathrm{E}}$ and the bispectrum $B_{\delta\delta \mathrm{E}}^\mathrm{equi}$ with $k_\mathrm{max}=2 \ h \, \mathrm{Mpc}^{-1}.$ Also shown are the mean $A_\mathrm{IA}$ for all masses, calculated as the average over mass bins weighted by the number of halos per bin.  For both power spectra \mbox{$k_\mathrm{min}\approx 0.1  \ h \, \mathrm{Mpc}^{-1}$} and for the bispectrum  \mbox{$k_\mathrm{min}\approx 0.7  \ h \, \mathrm{Mpc}^{-1}$}.   }
    \label{tab:A_IA_summary_mass}
  \centering
    \begin{tabular}{lccccccc}

\hline
\multicolumn{1}{l}{}&\multicolumn{1}{c}{}&\multicolumn{4}{c} {Subhalo mass range $[h^{-1}\mathrm{M}_\odot]$}&\multicolumn{1}{c}{}\\
\multicolumn{1}{l}{}&\multicolumn{1}{c}{Redshift} &\multicolumn{1}{r} {$10^{10}-10^{11}$} & \multicolumn{1}{r}{ $10^{11}-10^{12}$}& \multicolumn{1}{r}{ $10^{12}-10^{13}$}& \multicolumn{1}{r}{ $10^{13}-10^{14}$}&\multicolumn{1}{c}{Slope $\beta$}&\multicolumn{1}{c}{Mean $A_\mathrm{IA}$}\\
\hline
{From power spectrum $P_{\delta \mathrm{E}}$}&  0.0&0.74$\pm$0.01 &0.86$\pm$ 0.01&2.40$\pm$ 0.03&11.13$\pm$0.12&\ \ 0.392$\pm$0.002&0.91$\pm$0.01 \\

{}&0.5&1.02$\pm$0.01&1.31$\pm$0.01&4.26$\pm$0.04&19.38$\pm$0.14&\ \ 0.432$\pm$0.001&1.36$\pm$0.01\\

{}&1.0&1.59$\pm$0.01&2.10$\pm$0.02&6.68$\pm$0.06&28.87$\pm$0.21&\ \ 0.427$\pm$0.001&2.13$\pm$0.02\\\\

{From power spectrum $P_{\mathrm{E} \mathrm{E}}$} &0.0&1.40$\pm$0.02 &1.61$\pm$ 0.02&6.84$\pm$ 0.09&45.08$\pm$0.81&\ \ 0.507$\pm$0.002 &1.90$\pm$0.02\\

{}&0.5&1.96$\pm$0.02&2.25$\pm$0.02&9.14$\pm$0.09&63.43$\pm$1.01&\ \ 0.507$\pm$0.001&2.64$\pm$0.02\\

{}&1.0&2.69$\pm$0.02&3.21$\pm$0.02&13.54$\pm$0.15&112.53$\pm$1.47&\ \ 0.543$\pm$0.001&3.81$\pm$0.03\\\\

{From bispectrum $B_{\delta\delta\mathrm{E}}$}&0.0&0.94$\pm$0.01&0.99$\pm$0.01&2.33$\pm$0.03&9.74$\pm$0.16&\ \ 0.332$\pm$0.003&1.06$\pm$0.01\\

{}&0.5&1.07$\pm$0.01&1.25$\pm$0.01&3.79$\pm$0.05&17.59$\pm$0.23&\ \ 0.407$\pm$0.002&1.34$\pm$0.01\\

{}&1.0&1.58$\pm$0.02&2.07$\pm$0.02&5.94$\pm$0.09&33.88$\pm$0.39&\ \ 0.446$\pm$0.002&2.11$\pm$0.03\\
\hline
  \end{tabular}
 \end{table*}

\subsection{Analytical predictions for isosceles bispectra}\label{sect:isos_BS}
 It is worth exploring whether our analytical model can also be  applied to non-equilateral  bispectra since these can contain more information than equilateral configurations (Appendix \ref{sect:StoN}).  In this case it is not possible to obtain simplified expressions similar to those in equations~(\ref{eq:BSequiDDE}) to (\ref{eq:BSequiEEE}) because the perturbation theory kernels do not cancel out.  Instead we consider whether the intrinsic alignment amplitudes predicted from $B^{\mathrm{equi}}_{\delta\delta\mathrm{E} }$ are also valid for non-equilateral bispectra.  For illustration we again consider isosceles triangles with one side with magnitude $k$ and two sides with magnitude $2k$.

\begin{figure*}
\includegraphics[width=15cm]{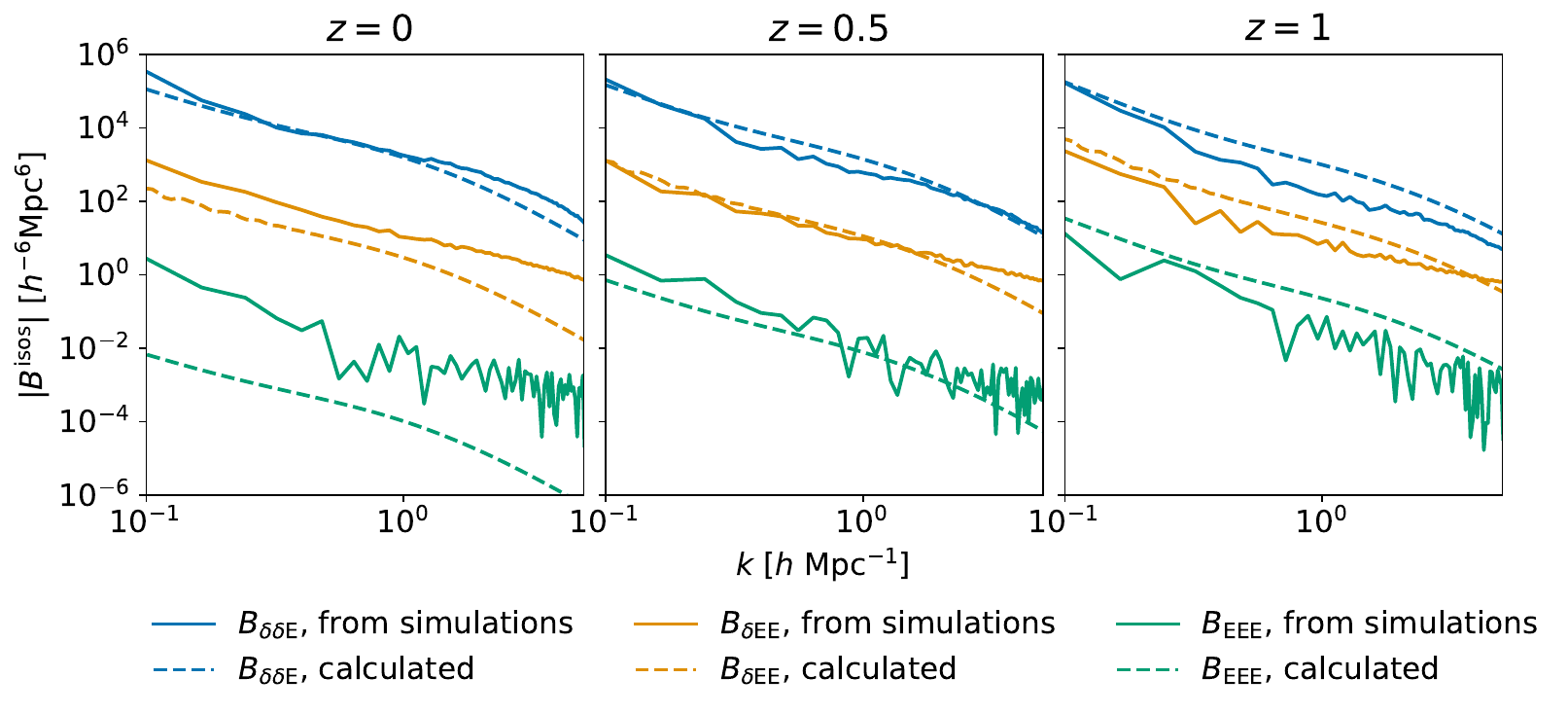}
\caption{ Intrinsic alignment bispectra for isosceles triangles with one side of magnitude $k$ and two sides of magnitude $2k$. \textit{Solid lines}:  Measured from simulations. \textit{Dashed lines}: Calculated from  equations~(\ref{eq:BSDDELA}) to  (\ref{eq:BSEEELA})  with best-fitting values of $A_{\mathrm{IA}}$ estimated from $B^\mathrm{equi}_{\delta \delta \mathrm{E}}$.  } \label{fig:isos_comp}
\end{figure*}

We take the estimates of $A_\mathrm{IA}$ obtained from $B^\mathrm{equi}_{\delta \delta \mathrm{E}}$ in the final column of Table~\ref{tab:A_IA_summary_mass}, noting that these are very similar to the estimates obtained from $P_{\delta \mathrm{E}}$. We calculate $f_\mathrm{IA}$ from equation (\ref{eq:fIA}) using the same assumptions as in Section~ \ref{sect:IAPS}, obtain  the non-linear matter power spectrum from Halofit \citep{takahashi2012revising}, and insert the resulting  values into equations~(\ref{eq:BSDDELA}) to  (\ref{eq:BSEEELA}).  In Fig.~\ref{fig:isos_comp} we compare these  predicted bispectra with those measured from simulations, finding a reasonable fit across scales and redshifts. This confirms that our empirical IA bispectrum model works for both equilateral and isosceles triangle configurations.

\subsection{Decorrelation of matter overdensity and E-mode fields }
In Appendix~\ref{sect:adjust} we discuss a possible phenomenological change to our two-point and three-point analytical models which adjusts them for the correlation between the matter overdensity field and the E-mode \lq field\rq.  We measure the correlation between the fields  from simulations, and introduce the resulting correlation coefficient into the analytical model. In most cases this improves the fit of the models considerably.  This is not central to our main results but may be worth considering for future intrinsic alignment modelling.

\section{Comparison with previous work}\label{sect:comp}
The most relevant previous work is by \citet{kurita2021power} who measured intrinsic alignment power spectra from simulations;  \citet{semboloni2008sources} who measured three-point intrinsic alignment statistics from simulations and compared them with cosmic shear; and \citet{pyne2021self} who used analytical models to investigate the effect of intrinsic alignments on the weak lensing power spectrum and bispectrum. 

Our present work is most directly comparable with \citet{kurita2021power} who used the same methodology but a different simulation suite.  Both studies explore  the correlation between the intrinsic shear field and  the matter overdensity field, and both consider dark matter halos rather than galaxies. The power spectrum results presented in the current work are consistent with those in \citet{kurita2021power}, giving confidence that the bispectrum measurements we present here are also sound.

 \citet{semboloni2008sources} built on the work of \citet{heymans2006potential} who measured two-point (GI) correlations between the shapes of foreground galaxies and the weak lensing shear of source galaxies in a suite of \textit{N}-body simulations, and also  shape--shape (II) correlations.  \citet{semboloni2008sources} used the same simulations to measure three-point statistics (GGG, GGI, GII and III correlations).   Their main findings were that for surveys whose median redshift was around 0.7 the II and III terms were consistent with zero, but that for shallower surveys with $z_\mathrm{med}\approx 0.3$ the II/GG and III/GGG ratios were non-zero, and the III ratio could be a factor of 10 higher than the II ratio. Later \citet{semboloni2010weak} also used these simulations to validate measurements of three-point shear statistics from observations. These studies are relevant to the present work in that they measured three-point IA statistics, but not easily comparable because they focused on IA contamination of the cosmic shear signal.

A likely reason for any inconsistencies between the findings in this paper and those of \citet{semboloni2008sources} is the different mass resolution of the simulations. The simulations used in \citet{semboloni2008sources} and related works were state of the art at the time, with $512^3$ particles in a periodic cubic box measuring $300 \ h \, \mathrm{Mpc}^{-1}$ per side. However the mass resolution was low with a particle mass of \mbox{$1.7\times 10^{10} \ h^{-1} \mathrm{M}_\odot$}, compared with \mbox{$4\times 10^7 \ h^{-1} \mathrm{M}_\odot$} in IllustrisTNG300.  As a result the smallest bound halos which \citet{heymans2006potential}  could  identify had masses several times $10^{11}  \ h^{-1} \mathrm{M}_\odot$ whereas in the current work we use halos with mass as low as around $4 \times 10^{10}\ h^{-1} \mathrm{M}_\odot$. This is despite the fact that we include only halos with at least 1000 particles, in contrast to  \citet{heymans2006potential} who allowed a considerably smaller minimum particle number.  \citet{semboloni2008sources} explicitly stated that the lack of low-mass halos in their sample was a limitation.
The results in \citet{heymans2006potential} and \citet{semboloni2008sources} also depend on models which they used to populate each halo with a single spiral or elliptical galaxy.
Although it would be possible to do an approximate lensing calculation to compare our results with \citet{semboloni2008sources}, in view of the many differences between the two studies we consider this would involve too many assumptions to be useful.

\citet{pyne2021self} modelled the GGI, GII and III correlations analytically using the same approach as in the present paper (the NLA model extended to three-point statistics). This work also found that intrinsic alignments affected two-point and three-point weak lensing statistics differently, although not consistently with the results in \citet{semboloni2008sources}. Again this disagreement may be attributable to the low mass resolution of their  simulations.

\section{Summary and discussion}\label{sect:discussion}

We have measured  intrinsic alignment bispectra of dark matter subhalos from the IllustrisTNG300-1 cosmological simulation suite, building on the power spectrum methodology developed by  \citet{kurita2021power}. We also measured  the IA power spectra $P_{\delta \mathrm{E}}$ and $P_{ \mathrm{E}\mathrm{E}}$ and confirmed that they are consistent with results obtained by \citet{kurita2021power} using the  DarkQuest simulation suite.

At all redshifts the cumulative signal-to-noise ratios of the IA bispectra were well below those we obtained for power spectra -- for example, \mbox{10-15 per cent} at $k_\mathrm{max}\approx 4 h \, \mathrm{Mpc}^{-1}$ compared with around 40 per cent for power spectra. 
The bispectrum $B_{\mathrm{E}\mathrm{E}\mathrm{E}}$ and measured bispectra involving B-modes have very low signal-to-noise ratios and provide no useful information.  However we found that the E-mode bispectra $B_{\delta\delta\mathrm{E}}$ and $B_{\delta\mathrm{E}\mathrm{E}}$ do have useful information content.  Signal-to-noise ratios for the non-equilateral triangles which we studied are notably higher than for equilateral triangles, suggesting that the common simplification of using only equilateral triangles in bispectrum analyses is sub-optimal \citep{yankelevich2022halo}.

All the IA power spectra and bispectra we studied showed similar strong relationships with subhalo mass. This can be traced back to the relationship between halo ellipticity and mass.  As discussed in Section~\ref{sect:ell_mass} this mass dependence has been noted by others in both simulations and observations.  Its origin is debated. \citet{smith2005triaxial} attributed it to higher mass halos forming later so they have had less time to virialise and therefore retain more memory of the tidal fields at the time they formed.  More recently \citet{xia2017halo} postulated that the IA strength depends independently on both halo formation time and mass, whereas \citet{piras2018mass} suggested that higher-mass halos experience stronger tidal fluctuations.  

We used the standard non-linear alignment model to estimate the intrinsic alignment amplitude from both $P_{\delta \mathrm{E}}$ and $P_{ \mathrm{E}\mathrm{E}}$, and found that our estimates are consistent with corresponding results obtained by \citet{kurita2021power}. This validated our two-point modelling.

For IA bispectra we used the analytical model from  \citet{pyne2021self}, which is in the spirit of the NLA model. From this we again estimated the intrinsic alignment amplitude, this time from equilateral bispectra.  We found that the best-fitting amplitudes  $A_\mathrm{IA}$ obtained using $B_{\delta\delta\mathrm{E}}^\mathrm{equi}$ were extremely close to those obtained from $P_{\delta \mathrm{E}}$.  It is not possible to use the same methodology to estimate $A_\mathrm{IA}$ from non-equilateral triangles but we showed that  the predicted $A_\mathrm{IA}$ from $B^\mathrm{equi}_{\delta\delta\mathrm{E}}$  produced an acceptable fit to simulation measurements of IA bispectra of specific isosceles triangles.

We fitted power-law relationships between the estimated  $A_\mathrm{IA}$ and subhalo mass,  obtaining almost identical relationships from $P_{\delta\mathrm{E}}$ and $B_{\delta\delta\mathrm{E}}^\mathrm{equi}$. These also agreed approximately with the relationship found by \citet{piras2018mass} using power spectrum measurements from the Millennium simulation.
It is interesting that all our estimated IA amplitudes are similar for the two lowest-mass bins and larger for the high-mass bins. This hints at a possible broken power-law relationship with mass similar to the luminosity relationship found by \citet{fortuna2021kids} for a sample of luminous red galaxies from KiDS-1000 \citep{kuijken2019fourth}. 

Our bispectrum measurements are an advance on the early three-point IA measurements from simulations reported in \citet{semboloni2008sources} since we have been able to take advantage of the improved resolution and methodology of the IllustrisTNG simulation suite. \citet{semboloni2008sources} focused on the magnitude of the IA effect relative to cosmic shear, whereas we considered the correlation between the IA shear and the matter overdensity field. There are also several other detailed differences between the two studies and, as discussed in Section \ref{sect:comp}, many assumptions would need to be made to compare our results directly with theirs.  It is more useful to consider the  results from  \citet{pyne2021self} who estimated ratios between IA and lensing signals using the same analytical models as in this paper. Like \citet{semboloni2008sources}, this work found that  two-point and three-point weak lensing statistics are affected differently by IA.  The present work has shown that our analytical models agree well with measurements from IllustrisTNG which validates the results reported in \citet{pyne2021self}. 

The fits between the measured and modelled IA power spectra and bispectra are not perfect. In  Appendix~\ref{sect:adjust} we suggest a possible phenomenological  modification of the analytical models based on the correlation between the matter overdensity and E-mode fields measured in simulations. This may be worth considering for future IA modelling.

Overall our results demonstrate that a single physically motivated analytical approach can be applied to both  two-point and three-point IA statistics, enabling a cleaner separation between IA and weak lensing signals. This opens up the prospect of using such a model in  joint  power spectrum--bispectrum analysis. \citet{pyne2021self} showed that such analysis can allow self-calibration  to mitigate intrinsic alignment contamination of weak lensing data in forthcoming surveys. 
This is  particularly pertinent in the light of advances in the measurement of three-point statistics from survey data.  \citet{secco2022darkb} recently reported  high signal-to-noise  detections of three-point shear correlations and aperture mass statistics in the first three years of data from the Dark Energy Survey. These measurements lay the foundations for joint two- and three-point cosmological analyses which will of course need tight control of systematics such as intrinsic alignments. 

This work has considered only dark matter halos and so has only limited application to observational data.  In future work we plan to build on the measurement techniques and modelling in this paper to confirm that the three-point approach can be extended to galaxies and to investigate the impact of galaxy characteristics and environment. 

\section*{Acknowledgements}
We thank the IllustrisTNG team for making their simulation data publicly available. We are also grateful to our anonymous reviewer for many constructive suggestions for improving the paper.
SP also thanks Harry Johnston and Anik Halder for helpful discussions.
This work was partially enabled by funding from the UCL Cosmoparticle Initiative.

\section*{Data Availability}
The data underlying this article will be shared on reasonable request to the corresponding author.

\bibliographystyle{mnras}
\bibliography{refs} 

\appendix
 
\section{Comparison between simulations and analytical calculations of the matter bispectrum}\label{sect:BHF_comp}
\begin{figure*}
\hspace{-3.5cm}
 \begin{subfigure}[b]{0.15\linewidth}
          \includegraphics[width=7cm]{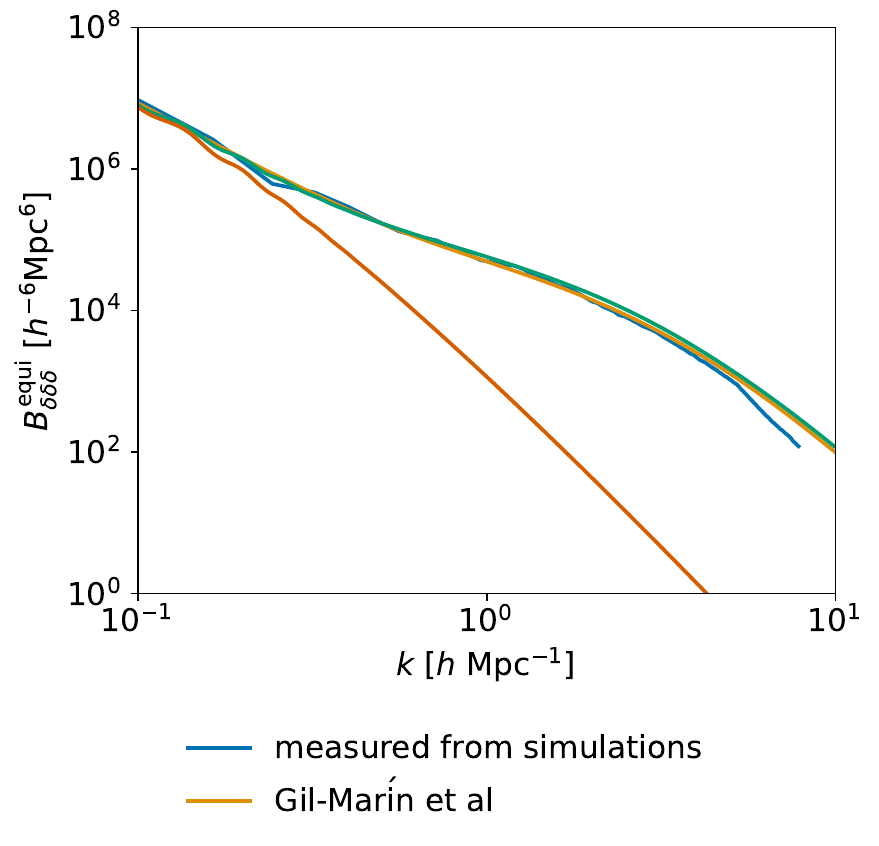}      
    \end{subfigure}
    \hspace{4.5cm}
    \begin{subfigure}[b]{0.15\linewidth}
          \includegraphics[width=7cm]{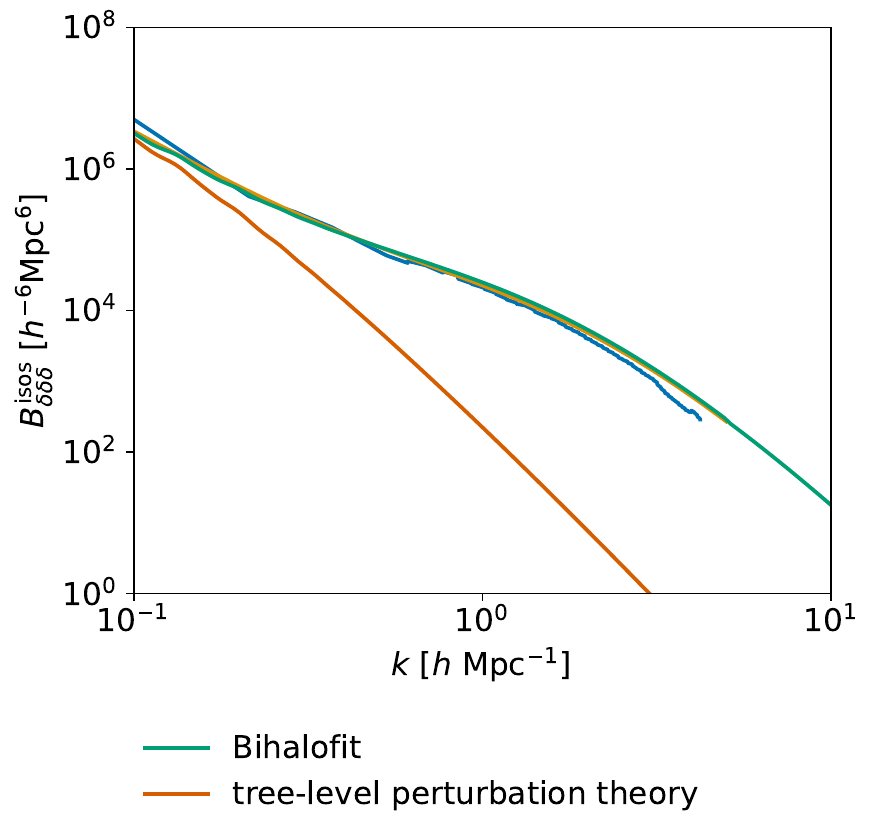}      
    \end{subfigure}
  \caption{ Comparison between the matter bispectrum at $z=1$ measured from the IllustrisTNG300-1 simulation, calculated using the fitting formula from \citet{gil2012improved}, and  calculated using the more accurate Bihalofit fitting formula \citep{takahashi2020fitting}.  Also shown are the bispectra calculated from tree-level perturbation theory \citep{bernardeau2002large}. \textit{Left}: Equilateral triangles, \textit{Right}: Isosceles triangles with sides in the ratio 2:2:1.}\label{fig:BScomp} 
\end{figure*}

To confirm the suitability of the perturbation theory-based formula from \citet{gil2012improved} for our purposes, Fig.~\ref{fig:BScomp} compares it with our matter bispectrum measurements  at $z=1$ from the IllustrisTNG300-1 simulations.   Also shown are more recent analytical estimates using Bihalofit \citep{takahashi2020fitting} as well as  tree-level perturbation theory bispectra \citep{bernardeau2002large}. We show results for equilateral triangles and for isosceles triangles with sides in the ratio 2:2:1.   All the separate results for the non-linear bispectrum are consistent at the scales which we  are interested in, with the simulation measurements and analytical estimates from \citet{gil2012improved} only diverging significantly for $k \gtrsim 3 h \, \mathrm{Mpc}^{-1}$.

\section{Signal-to-noise ratios}\label{sect:StoN}

We approximate the cumulative signal-to-noise ratio (SNR) of a power spectrum $P(k)$ as
\begin{align}
\Bigg(\frac{\mathrm{S}}{\mathrm{N}} \Bigg)^2&= \sum_{k_i=k_\mathrm{min}}^{k_\mathrm{max}} \frac{[P(k_i)]^2}{ \sigma^2_i} \label{eq:SNR}\ ,
\end{align}
where $\sigma^2_i$ is the variance in the $i$th bin, measured using jackknife sampling, and we consider only diagonal  terms of the covariance. Similar definitions apply to the IA power spectra and bispectra.  Equation (\ref{eq:SNR}) is a  simplification which somewhat overestimates the signal-to-noise ratio. Nevertheless it allows a useful comparison between the information content of different spectra.

\begin{figure*}
\centering

 \includegraphics[width=15cm]{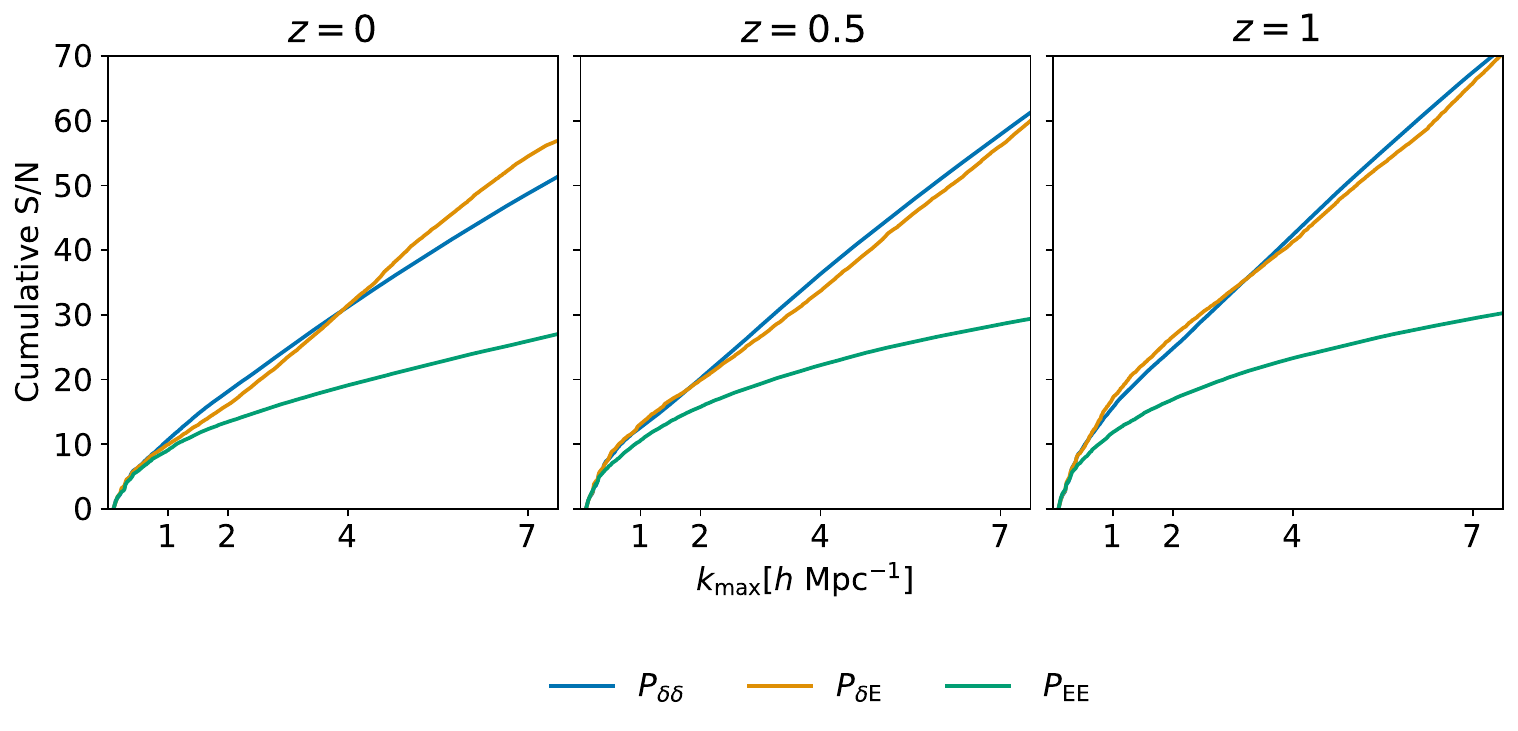}  
\caption{ Cumulative signal-to-noise ratios for the matter power spectrum and E-mode intrinsic alignment power spectra measured from simulations. }\label{fig:SNR_PS}

\end{figure*}

\begin{figure*}
\centering

\centering

 \includegraphics[width=15cm]{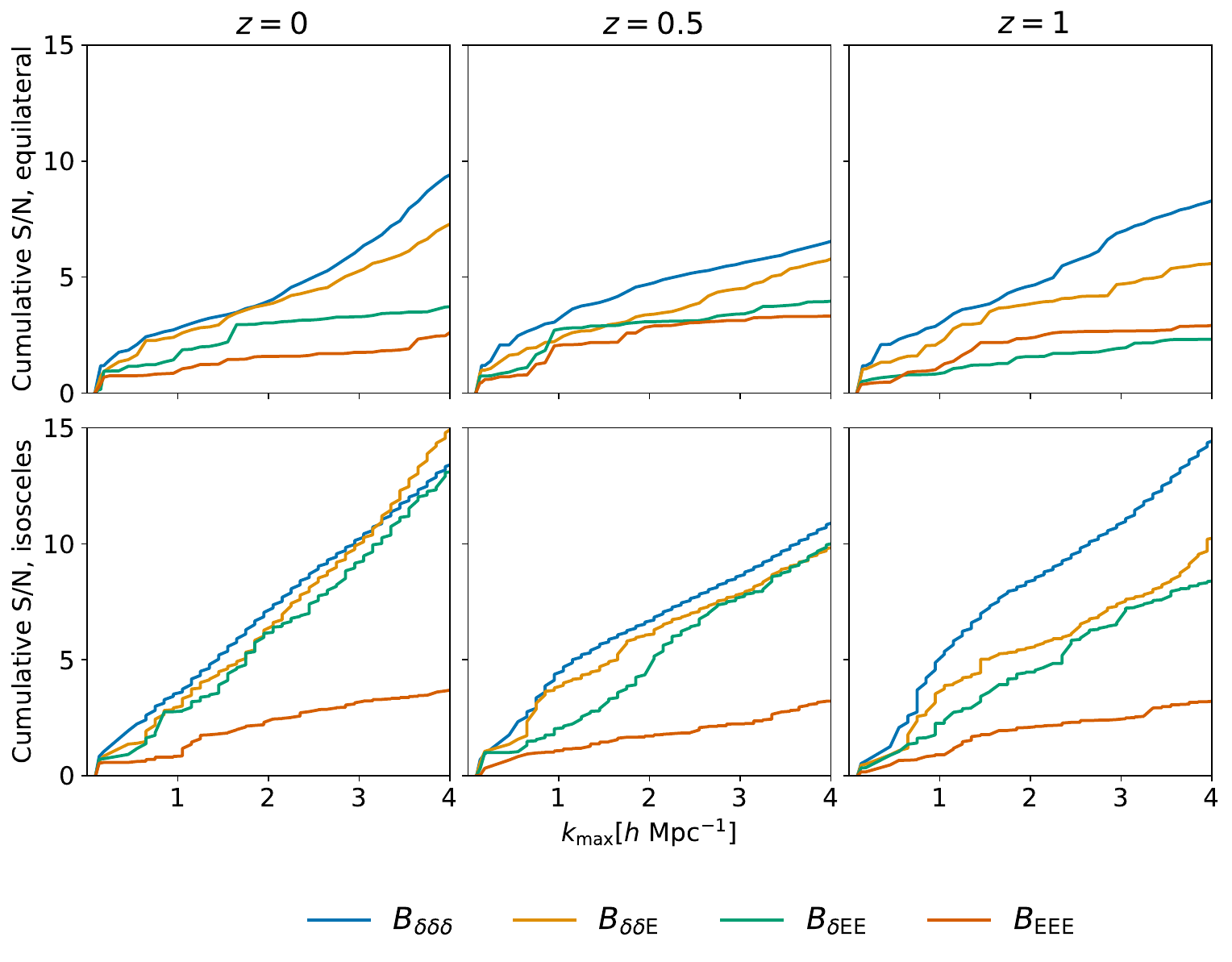}  
\caption{ Cumulative signal-to-noise ratios for the matter bispectrum and E-mode intrinsic alignment bispectra measured from simulations.  \textit{Top}: Equilateral triangles. \textit{Bottom}: Isosceles triangles with sides in ratio 2:2:1.}\label{fig:SNR_BS_E}
\end{figure*}

Figure \ref{fig:SNR_PS} shows, as a function of $k_\mathrm{max}$, the cumulative SNR for the matter power spectrum  and also for E-mode IA power spectra.  
Figure~ \ref{fig:SNR_BS_E}  shows similar information for the E-mode intrinsic alignment bispectra, for  equilateral triangles and for isosceles triangles with sides in the ratio 2:2:1.

The main messages to take from these figures are that the SNR for $P_{\mathrm{E}\mathrm{E}}$ is much weaker than for $P_{\delta\mathrm{E}}$; the power spectra contain much more information than the bispectra; and the intrinsic alignment spectra $B_{\delta\mathrm{E}\mathrm{E}}$ and $B_{\mathrm{E}\mathrm{E}\mathrm{E}}$ contain very little signal in the IllustrisTNG300-1 volume. For this reason we use only $B_{\delta\delta\mathrm{E}}$ in most parts of this work.  The $P_{\delta\mathrm{E}}$ SNR appears surprisingly strong in comparison to that for $P_{\delta\delta}$ but this could be due to the simplifications in our calculations.  We also confirmed that in each case the cumulative SNRs for individual mass bins are typically within 80 per cent of the SNRs for the whole sample.

\section{Phenomenological adjustment to analytical models}\label{sect:adjust}

\begin{figure*}
\includegraphics[width=15cm]{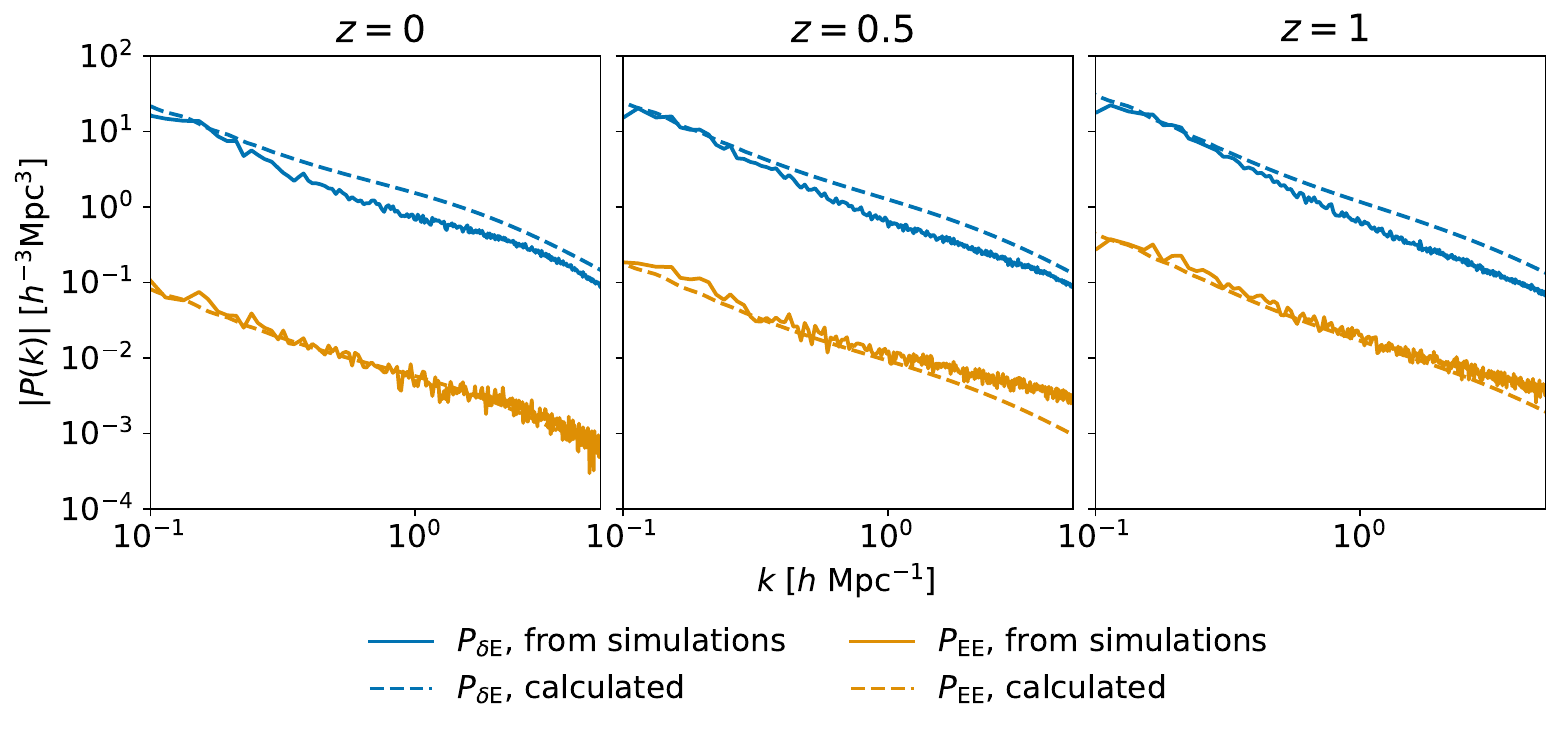}
\caption{ Intrinsic alignment power spectra $P_{\delta \mathrm{E}}$ and $P_{\mathrm{E}\mathrm{E}}$ estimated from simulations (solid lines) and  calculated from the non-linear alignment model with best-fitting values of $A_{\mathrm{IA}}$ estimated from $P_{\delta \mathrm{E}}$ (dashed lines).  } \label{fig:PS_comp}
\end{figure*}

\begin{figure*}
\includegraphics[width=15cm]{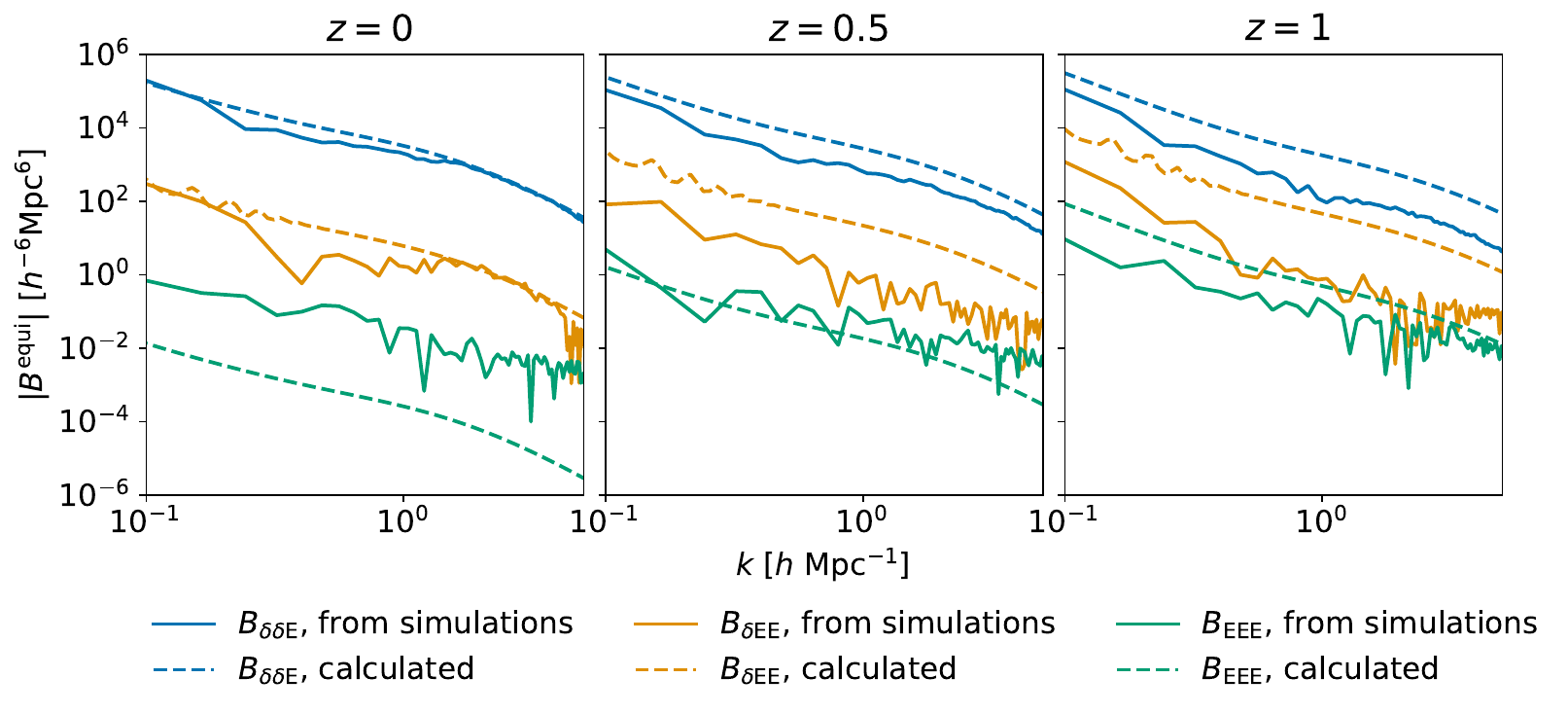}
\caption{ Intrinsic alignment bispectra for equilateral triangles. \textit{Solid lines}:  Measured from simulations. \textit{Dashed lines}: Calculated from  equations~(\ref{eq:BSDDELA}) to (\ref{eq:BSEEELA}) \, with best-fitting values of $A_{\mathrm{IA}}$ estimated from $B^\mathrm{equi}_{\delta \delta \mathrm{E}}$.  }  \label{fig:equi_comp}
\end{figure*}

In Fig.~\ref{fig:isos_comp} we compare IA isosceles bispectra measurements from simulations with analytical predictions and show that there is a close but not perfect match between the two estimates.  
Figures~\ref{fig:PS_comp}  and \ref{fig:equi_comp} respectively show analogous results for intrinsic alignment  power spectra and equilateral  bispectra.  Again the fits are close but not exact.

\begin{figure*}
 \centering 
 \hspace{0.3cm}
\includegraphics[width=15cm]{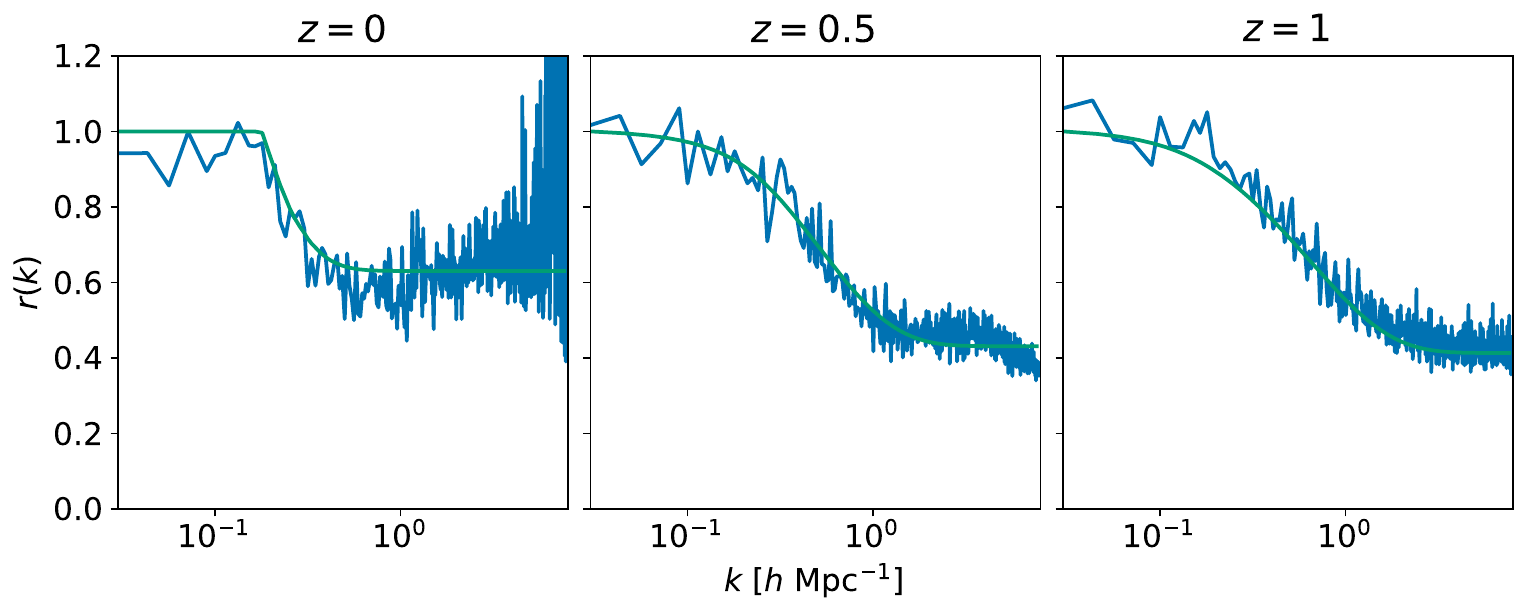} 
\caption{\textit{Blue}: Correlation coefficients $r(k)$  between the E-mode field and the matter density field as defined by equation~(\ref{eq:r}), for three redshifts.  \textit{Green}: Logistic curve fitted to the calculated correlation coefficients using equation (\ref{eq:logistic}) with parameter values from Table~\ref{tab:params}. }\label{fig:corrcoeff}
\end{figure*}

Looking more closely at  Fig.~\ref{fig:PS_comp}, the simulations and analytical calculations match well for $P_{\mathrm{E}\mathrm{E}}$; the discrepancy is only in $P_{\delta\mathrm{E}}$ which correlates the matter overdensity  and  E-mode fields (the latter is not a true field but we treat it as such).  From Fig.~\ref{fig:isos_comp} and Fig.~\ref{fig:equi_comp} we see that the fit is not perfect for any of the bispectra. The measured $B_{\mathrm{E}\mathrm{E}\mathrm{E}}$ (isosceles and equilateral) are  noisy and in this case part of the discrepancy may be due simply to the uncertainty in the measurements.  
We therefore consider here how to improve the fits for the spectra which depend on both the matter density and  E-mode fields. We tackle this by considering the correlation between the measured power spectra of the two fields.  The correlation coefficient $r(k)$ between the power spectra can be defined as   \citep{kurita2021power} 
\begin{align}
r^2(k) &= \frac{6}{5}  \frac{P_{\delta\mathrm{E}}(k)^2}{P_{\delta\delta}(k)P_{\mathrm{E}\mathrm{E}}(k)}\ . \label{eq:r}
\end{align}
The normalizing factor 6/5 arises from integration over $\mu$ of the terms in $(1-\mu^2)$ in equations~(\ref{eq:PSDELA}) and (\ref{eq:PSEELA}):
\begin{align}
\frac{\big[ \int_0^1 (1-\mu^2)  \, \mathrm{d}\mu\big]^2}{ \int_0^1 (1-\mu^2)^2\, \mathrm{d}\mu}
= \frac{5}{6}\ .
\end{align}
 With this normalization, $r(k)$ should be equal to unity if the non-linear alignment model is valid. Our calculated correlation coefficients at three redshifts are shown in blue in Fig.~\ref{fig:corrcoeff}.  The  coefficients asymptote to approximately one as $k\to0$ but reduce to a lower, redshift-dependent value for $k\gtrsim 1$, apart from  $z=0$ where the correlation coefficient is very noisy at large values of $k$. 
 
To model the correlation coefficients at each redshift we fit a generalized logistic curve of the form
\begin{align}
 r(k)&= r_0 + \frac{r_1-r_0}{\big[1+a\exp{(-b(k-k_0)\big]}^{1/\nu}} \ .\label{eq:logistic}
\end{align}
This has six free parameters: $r_0$ and $r_1$ are the lower and upper asymptotes of the curve, $k_0$ determines the $k$ value at which the curve starts to decline, and the remaining parameters, $a$, $b$ and $\nu$, control the shape of the curve. In particular, $b$ controls the rate at which the curve decreases as $k$ increases.  We normalize the fitted curves to be equal to one at $k_\mathrm{min}$. For  $z=0$, because of the noisy data at larger $k$,  we set $r(k)=r(1.0)$ for $k\ge1.0\, h \mathrm{Mpc}^{-1}$.   The fitted curves are shown in green in Fig.~\ref{fig:corrcoeff}.  

 It is difficult to fit a single model which takes redshift into account since we only have data for three redshifts. Instead we fit the model separately for each redshift, leading to  the best-fitting parameter values shown in  Table~\ref{tab:params}. The values for $z=0.5$ and $z=1$ are similar so it is plausible that  single model could in fact be constructed for both these redshifts. 
 \begin{table}
\vspace{0.9cm}
\caption{Best-fitting parameter values and normalization factor for the modelled correlation coefficients given by equation~(\ref{eq:logistic}).}
    \label{tab:params}
  \centering
    \begin{tabular}{cccccccc}
\hline
    Redshift&$r_0$&$r_1$&$a$&$b$&$\nu$&$k_0$&Norm\\\hline
0.0&0.6 &0.95&1.03&614&57.8&0.19&0.95 \\
0.5 & 0.43&1.02&1.72&12.5&6.00&0.16 &1.00\\
 1.0 & 0.43 & 1.06&2.20&19.9&13.0&0.08&1.04 \\
 \hline  
   \end{tabular}
    \end{table}

We now introduce the correlation coefficient into the two-point analytical model by rewriting equation~(\ref{eq:PSDELA}) as
\begin{align}
P_{\delta\mathrm{E}}(k)&= r(k)f_{\mathrm{IA}}P_\mathrm{NL}(k)\ ,\label{eq:PSDELA1}
\end{align}
for each redshift. Thus the model now includes the simulation-based correlation between the fields.
Similarly we incorporate correlation coefficients into the three-point model given by equations~(\ref{eq:BSDDELA})  and  (\ref{eq:BSDEELA}).  So, since $P_{\mathrm{EE}}$ is unchanged, we have
\begin{align}
B_{\delta\delta  \mathrm{E}}(\bm{k}_1,\bm{k}_2,\bm{k}_3)&=
 2\, \Big[ f_{\mathrm{IA}}^2F_2^{\mathrm{eff}}(\bm{k}_1,\bm{k}_2) P_\mathrm{NL}(k_1)P_\mathrm{NL}(k_2)\label{eq:BSDDELA1}\notag\\
 &\hspace{0.4cm}+ r(k_2)f_{\mathrm{IA}}F_2^{\mathrm{eff}}(\bm{k}_2,\bm{k}_3) P_\mathrm{NL}(k_2)P_\mathrm{NL}(k_3)\notag\\
&\hspace{0.4cm} +r(k_3)f_{\mathrm{IA}}F_2^{\mathrm{eff}}(\bm{k}_3,\bm{k}_1) P_\mathrm{NL}(k_3)P_\mathrm{NL}(k_1)\Big]\ ,
\end{align}
and similarly for $B_{\delta \mathrm{EE}}$.

\begin{figure*}
\includegraphics[width=15cm]{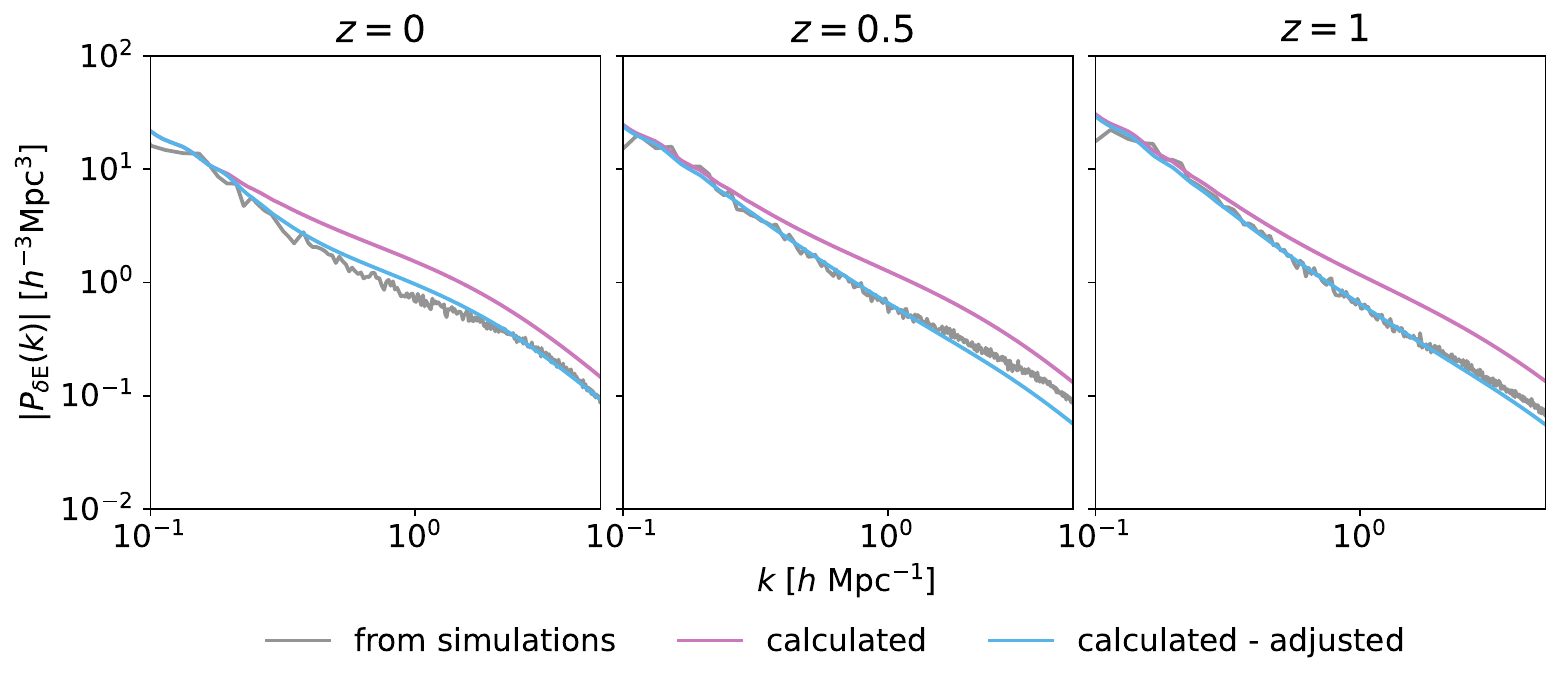}
\caption{ Intrinsic alignment power spectrum $P_{\delta \mathrm{E}}$ measured from simulations  and  calculated from the non-linear alignment model with and without adjustment by the correlation coefficient $r(k)$ given by equation (\ref{eq:r}). }\label{fig:PS_comp_r} 
\end{figure*}

Figure~\ref{fig:PS_comp_r}  compares the unadjusted and adjusted analytical results to the simulation measurements for $P_{\delta\mathrm{E}}$.  In this case, by construction, the phenomenological modification virtually eliminates the discrepancy between the simulations and analytical calculations.  

\begin{figure*}
\includegraphics[width=15cm]{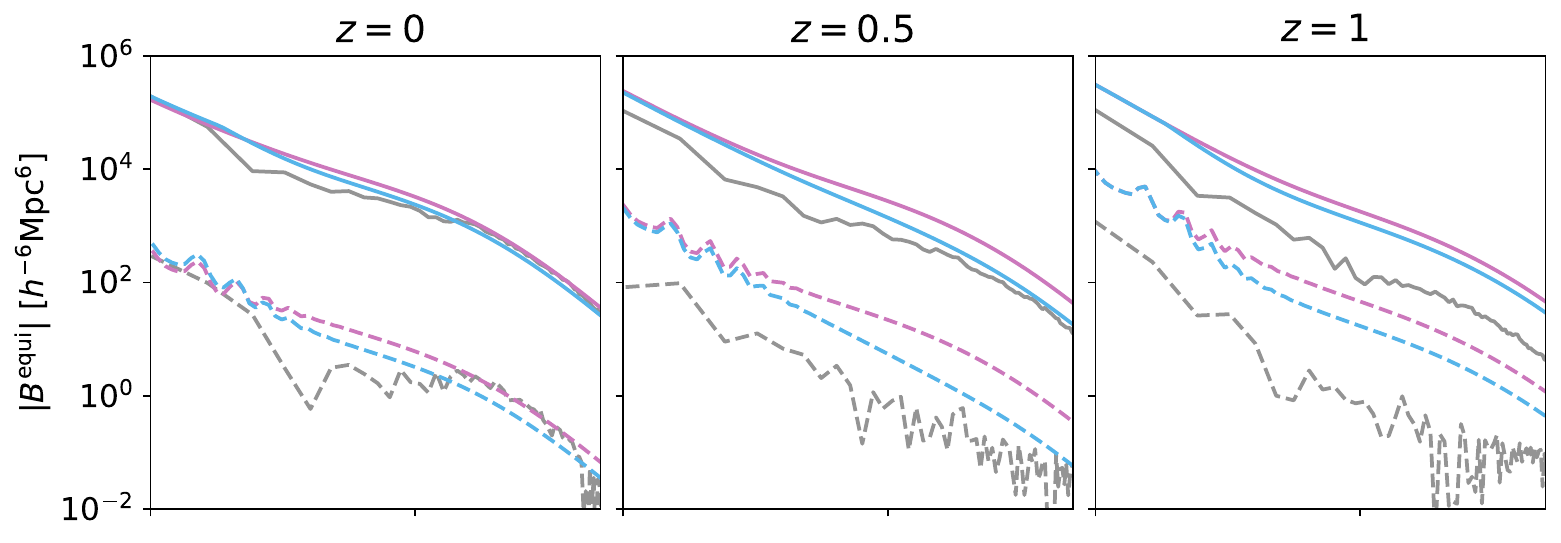}
\includegraphics[width=15cm]{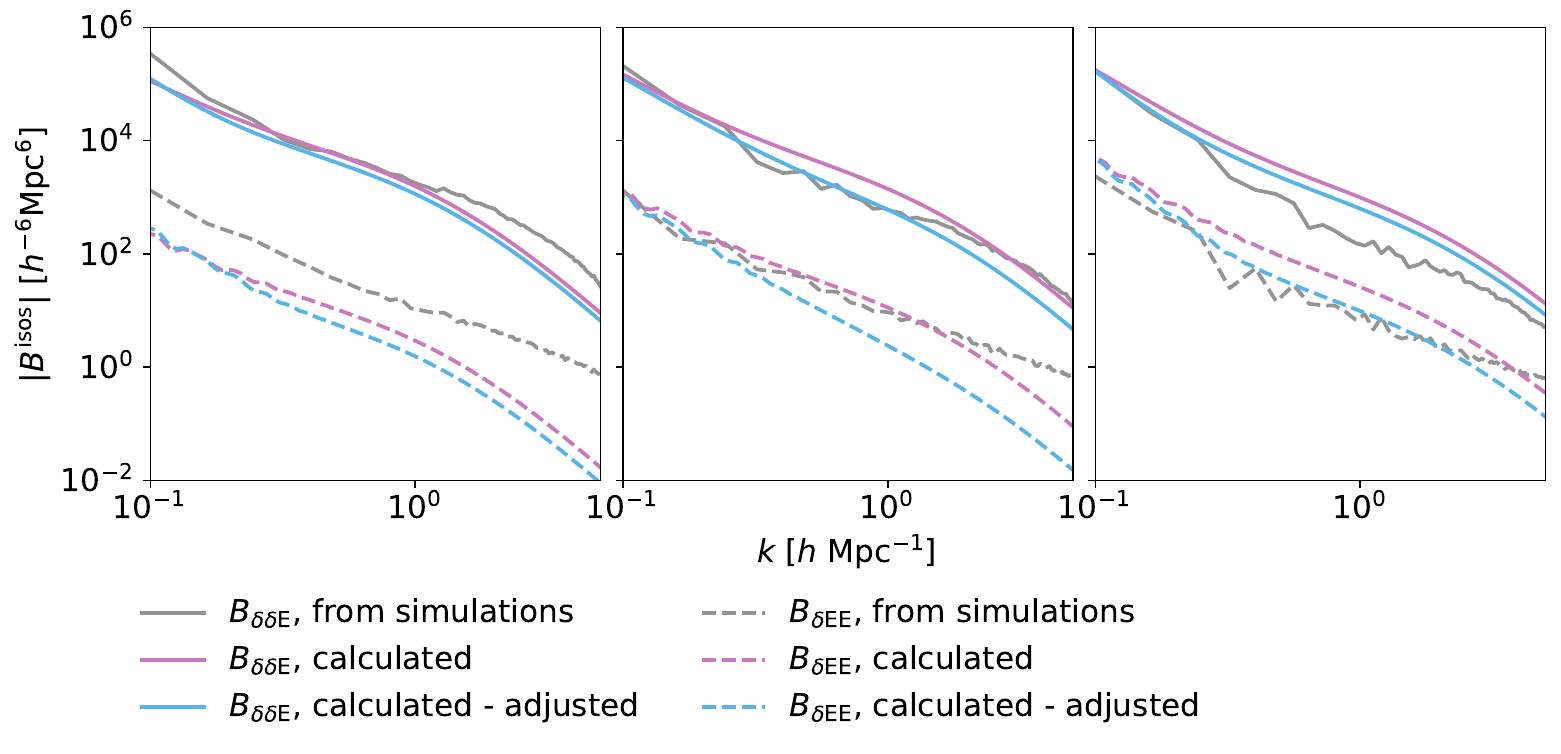}
\caption{ Intrinsic alignment bispectra  $B_{\delta\delta\mathrm{E}}$ and $B_{\delta\mathrm{E}\mathrm{E}}$ estimated from simulations and  calculated analytically with and without adjustment by the correlation coefficient $r(k)$ given by equation~(\ref{eq:r}).  \textit{Top}: Equilateral triangles, \textit{Bottom}: Isosceles triangles with sides in ratio 2:2:1. } \label{fig:isos_comp_r}
\end{figure*}

Figure~\ref{fig:isos_comp_r} shows similar results for the bispectra $B_{\delta\delta\mathrm{E}}$ and $B_{\delta\mathrm{E}\mathrm{E}}$ for equilateral and isosceles triangles.  Here the benefits of introducing the correlation coefficient are less clear, especially for  $B_{\delta\mathrm{E}\mathrm{E}}$ where neither version of the analytical model fits the simulations very well, especially at higher redshifts. Nevertheless we conclude that it may be worth considering this decorrelation between the density field and the intrinsic shape in future two-point and three-point IA models.

%
\bsp	
\label{lastpage}
\end{document}